\documentclass[notitlepage,superscriptaddress,showpacs,a4paper,nobalancelastpage,twocolumn,aps,prb,longbibliography]{revtex4-2}
\usepackage[english]{babel}
\RequirePackage[T1]{fontenc}
\RequirePackage{times} 
\usepackage[subpreambles=true]{standalone}

\usepackage{siunitx} 
\usepackage{amsfonts}
\usepackage{amsmath}
\usepackage{amssymb}
\usepackage{graphicx,epstopdf}
\usepackage{appendix}
\usepackage{xspace}
\usepackage{array}
\usepackage[hidelinks]{hyperref}
\usepackage{color}
\usepackage{xcolor}
\usepackage{bm}
\usepackage{verbatim}
\usepackage{tikz}
\usepackage{float}
\usepackage{appendix}
\usepackage{pdfpages}
\usepackage{lipsum}
\usepackage{booktabs}
\usepackage{tabularx}
\usepackage{longtable}
\usepackage{makecell}
\usepackage{threeparttable}
\allowdisplaybreaks[3]

\RequirePackage{cleveref}

  \AtBeginDocument{

\crefname{table}{Table}{Tables}
\crefname{figure}{Figure}{Figures}
\crefname{equation}{eq.}{eqs.}
\Crefname{equation}{Eq.}{Eqs.}
\crefname{section}{sect.}{sects.}
\Crefname{section}{Sect.}{Sects.}

\crefname{lemma}{lemma}{lemmas}
\Crefname{lemma}{Lemma}{Lemmas}
\crefname{theorem}{theorem}{theorems}
\Crefname{theorem}{Theorem}{Theorems}
\crefname{appendix}{appendix}{appendixs}
\Crefname{appendix}{Appendix}{Appendix}

\crefname{algorithm}{algorithm}{algorithms}
\Crefname{algorithm}{Algorithm}{Algorithms}
\crefname{assumption}{assumption}{assumptions}
\Crefname{assumption}{Assumption}{Assumptions}
\crefname{corollary}{corollary}{corollarys}
\Crefname{corollary}{Corollary}{Corollarys}
\crefname{definition}{definition}{definitions}
\Crefname{definition}{Definition}{Definitions}
\crefname{example}{example}{examples}
\Crefname{example}{Example}{Examples}
\crefname{problem}{problem}{problems}
\Crefname{problem}{Problem}{Problem}
\crefname{proposition}{proposition}{propositions}
\Crefname{proposition}{Proposition}{Propositions}
\crefname{remark}{remark}{remarks}
\Crefname{remark}{Remark}{Remarks}

 }

\renewenvironment{itemize}{
\begin{list}{$\bullet$}{
\labelwidth=4em
\labelsep=0.5em
\leftmargin=0em
\rightmargin=0em
\parsep=\parskip
\itemsep=0em
\topsep=0em
\itemindent=2.1em
}
}{\end{list}}

\makeatletter

\AtBeginDocument{\let\LS@rot\@undefined}
\makeatother
\usetikzlibrary{calc,3d}
\usetikzlibrary{arrows}
\usetikzlibrary{snakes}

\usepackage[normalem]{ulem}

{\par}
\newcommand{\add}[1]{{\color{black} #1}}
\newcommand\rmv{\bgroup\markoverwith {\textcolor{red}{\rule[0.5ex]{2pt}{0.4pt}}}\ULon}

\begin{document}
\title{Physical Neural Networks with Self-Learning Capabilities}

\author{Weichao Yu}
\affiliation{State Key Laboratory of Surface Physics and Institute for Nanoelectronic Devices and Quantum Computing, Fudan University, Shanghai 200433, China}
\affiliation{Zhangjiang Fudan International Innovation Center, Fudan University, Shanghai 201210, China}

\author{Hangwen Guo}
\email{hangwenguo@fudan.edu.cn}
\affiliation{State Key Laboratory of Surface Physics and Institute for Nanoelectronic Devices and Quantum Computing, Fudan University, Shanghai 200433, China}
\affiliation{Zhangjiang Fudan International Innovation Center, Fudan University, Shanghai 201210, China}
\affiliation{Shanghai Research Center for Quantum Sciences, Shanghai 201315, China}

\author{Jiang Xiao}
\email{xiaojiang@fudan.edu.cn}
\affiliation{State Key Laboratory of Surface Physics and Institute for Nanoelectronic Devices and Quantum Computing, Fudan University, Shanghai 200433, China}
\affiliation{Zhangjiang Fudan International Innovation Center, Fudan University, Shanghai 201210, China}
\affiliation{Department of Physics, Fudan University, Shanghai 200433, China}
\affiliation{Shanghai Research Center for Quantum Sciences, Shanghai 201315, China}
\affiliation{Collaborative Innovation Center of Advanced Microstructures, Nanjing 210093, China}

\author{Jian Shen}
\email{shenj5494@fudan.edu.cn}
\affiliation{State Key Laboratory of Surface Physics and Institute for Nanoelectronic Devices and Quantum Computing, Fudan University, Shanghai 200433, China}
\affiliation{Zhangjiang Fudan International Innovation Center, Fudan University, Shanghai 201210, China}
\affiliation{Department of Physics, Fudan University, Shanghai 200433, China}
\affiliation{Shanghai Research Center for Quantum Sciences, Shanghai 201315, China}
\affiliation{Collaborative Innovation Center of Advanced Microstructures, Nanjing 210093, China}

\begin{abstract}
Physical neural networks are artificial neural networks that mimic synapses and neurons using physical systems or materials. These networks harness the distinctive characteristics of physical systems to carry out computations effectively, potentially surpassing the constraints of conventional digital neural networks. A recent advancement known as ``physical self-learning'' aims to achieve learning through intrinsic physical processes rather than relying on external computations. This article offers a comprehensive review of the progress made in implementing physical self-learning across various physical systems. Prevailing learning strategies are discussed that contribute to the realization of physical self-learning. Despite challenges in understanding fundamental mechanism of learning, this work highlights the progress towards constructing intelligent hardware from the ground up, incorporating embedded self-organizing and self-adaptive dynamics in physical systems.
\end{abstract}

\maketitle

\section{Introduction}

Biological brains outperform state-of-the-art machine intelligence in many aspects especially in terms of energy efficiency. For example, it's estimated that the energy consumed by a modern supercomputer to train a natural language processing model is about \SI{1000}{kWh}, which is sufficient for human brains to operate for full six years \cite{markovic_physics_2020}.  

To emulate the energy-efficient computational capabilities of biological brains, the field of neuromorphic computing, or brain-inspired computing, has emerged. This computational paradigm traces its origins back to Carver Mead at Caltech in the 1980s \cite{furber_large-scale_2016}, where the development of bio-inspired microelectronic devices took place. The proposition was made that large-scale adaptive analog systems have the potential to learn about their environment while consuming significantly less power \footnote{Another Way Of Looking At Lee Sedol vs AlphaGo · Jacques
Mattheij}, thereby harnessing the full capabilities of silicon fabrication on a wafer scale \cite{mead_neuromorphic_1990}. Noteworthy examples of state-of-the-art large-scale neuromorphic systems include TrueNorth, Neurogrid, BrainScaleS, SpinNNaker, Loihi, Qualcomm, and Tianjic, among others \textit{etc.} \cite{jeong_memristors_2016,furber_large-scale_2016,pei_towards_2019,zhang_brain-inspired_2020}.

\begin{figure*}[t]
\centering
\includegraphics[width=1.8\columnwidth]{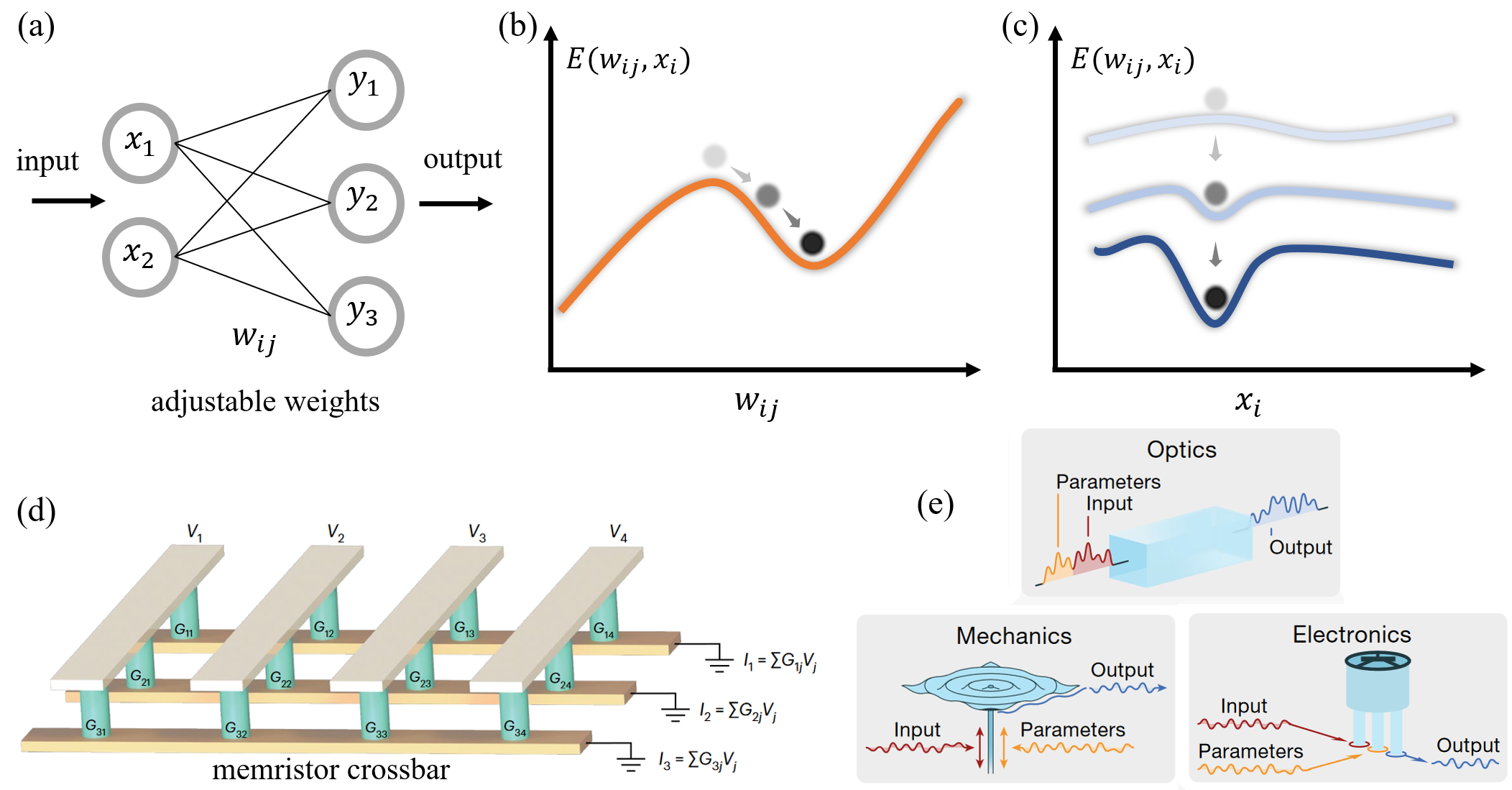}
\caption{(a) Structure of Artificial Neural Networks (ANN). Only one layer is shown for simplicity. Data are received and transmitted in a feedforward way through the network. (b)(c) When the neural network is learning the vector $x_i$, it's actually looking for an appropriate set of weights $w_{ij}$ that minimizes the loss function $E(w_{ij},x_i)$. Bias $\theta_i$ is not shown for simplicity. \add{(d) Memristor crossbar array is a type of PNN with isomorphic structure as ANN shown in (a). (e) Three types of PNNs which perform physical mapping from input to output, but not isomorphic to ANN shown in (a). The weights (and bias) of PNNs can be either internal physical degrees of freedom or external control parameters. The learning process of PNNs can be represented in the same way as demonstrated in (b) and (c). Reprinted with permission from \cite{wright_deep_2022,liang_physical_2024}.}}
\label{fig:1}
\end{figure*}

While the mechanism behind how the brain's vast number of neurons (approximately one hundred billion) and synapses (around one hundred trillion) achieve flexible cortical configurations remains as a fundamental question in neuroscience \cite{chialvo_emergent_2010}, computational scientists have developed machine learning models that attempt to mimic the operational principles of the human brain. One such model is the Artificial Neural Network (ANN). An ANN is a computational framework consisting of numerous interconnected elements, known as neurons, which are capable of receiving, processing, and transmitting information. The connections between neurons, referred to as synapses, are adjustable and possess the ability to adapt and learn. A typical ANN is structured into multiple layers, as depicted in Figure 1(a), where the value of each neuron, denoted as $y_j$, is determined by the collective input from all neurons, $x_i$, in the preceding layer \cite{rojas_neural_2013}
\begin{equation}
y_j=\mathcal{F}\left(\sum_{i}w_{ij}x_i+\theta_i\right),
\end{equation}
with $w_{ij}$ the adjustable synaptic weights describing how much the input neuron $x_i$ contribute and $\theta_i$ the bias locally applied on output neuron. The activation function $\mathcal{F}$ is nonlinear and can take various forms, such as Sigmoid, Tanh, and ReLU, \textit{etc}. There are two stages for a working ANN: (i) The learning process, where the network learns from a large amount of training data by adjusting weights as well as bias. The state of the network can be characterized by the loss function $E(w_{ij},x_i)$, and during the learning phase, the neural network essentially modulates the landscape of the loss function, as schemed in \cref{fig:1}(b,c). When the training is finished, the network moves to the next stage (ii) the inference process, where the network makes predictions on new data according to the weights determined in the first stage. With an increase in depth (number of layers), a neural network can express highly complex functions and learn more complex features from the training data \cite{mehta_high-bias_2019}. As a result, there is a widespread understanding that ``Multi-layer neural networks can do everything'' \cite{rojas_neural_2013}, including emulating physical phenomena and discovering new physical concepts \cite{iten_discovering_2020,gigli_predicting_2023}.

Although the ANN model has achieved remarkable success, it faces significant challenges, including high energy costs in both learning and inference processes, as well as a heavy reliance on external algorithms during learning. Enhancing computational models, such as ANN, is a critical objective in the field of Artificial Intelligence and machine learning. Neuromorphic computing, on the other hand, aims to address these challenges by mapping computational models (\textit{e.g.}, the one shown in \cref{fig:1}(a)) onto hardware through the design of physical neurons and synapses. State-of-the-art neuromorphic chips, primarily based on complementary metal-oxide-semiconductor (CMOS) technology, have achieved tremendous success and surpassed conventional processors like the graphic processing unit (GPU) \cite{zhang_neuro-inspired_2020}. Other physical platforms, including photonics \cite{sui_review_2020,shastri_photonics_2021,ashtiani_-chip_2022}, spintronics \cite{grollier_neuromorphic_2020,zhou_prospect_2021,zhang_spintronic_2020,cai_spintronics_2023}, superconducting Josephson junctions \cite{russek_stochastic_2016}, mechanical metamaterials \cite{pashine_directed_2019,stern_supervised_2020} and two-dimensional materials \cite{zhu_hybrid_2023}, among others \cite{del_valle_challenges_2018,zhao_reliability_2020}, are also making significant advancements.

This review focuses on a specific subgroup within neuromorphic computing known as Physical Neural Networks (PNNs), where computation and training occur within the physical processes of the aforementioned physical systems \cite{wright_deep_2022} (\cref{fig:1}(e)). \add{Both PNNs and CMOS neuromorphic chips can be considered as physical implementation of deep ANNs, which benefit from significant energy efficiency in the inference process due to physical acceleration. For instance, matrix multiplication can be efficiently performed in parallel either by leveraging Kirchhoff's law in electric circuits or by utilizing wave interference and scattering in photonic circuits \cite{khoram_nanophotonic_2019,christensen_2022_2022}. However, there is a fundamental distinction between PNNs and CMOS neuromorphic chips. While PNNs directly train the hardware's physical transformations to carry out desired computations, CMOS neuromorphic chips implement trained mathematical transformations by designing hardware with strict, operation-by-operation mathematical isomorphism \cite{wright_deep_2022}. Nowadays, most of the neuromorphic chips perform inference \textit{in situ} but are trained \textit{ex situ}, so that external computational resources are still required during the training process, which poses a bottleneck for further improving the speed and energy efficiency of neuromorphic computing.} \add{With their inherent capability of physical self-learning, PNNs offer a promising avenue for training without extensive reliance on external resources. Additionally, PNNs can harness the abundant internal degrees of freedom within physical systems, providing opportunities to enhance their capacity. However, several aspects of PNNs, including scalability, learning efficiency, noise tolerance, and the emergence of properties in complex systems beyond simple human-designed architectures, remain open questions that warrant further investigation. These areas present intriguing research directions to fully unlock the potential of PNNs and deepen our understanding of intelligent systems.} This review primarily focuses on the learning (training) process in PNNs, particularly exploring the emerging self-learning capabilities embedded within these physical systems.

\section{Physical Neural Networks}

Artificial Neural Networks (ANN) are virtual neural networks created in computers, with neurons and weights stored digitally. In contrast, Physical Neural Networks (PNN) have neurons and weights represented in physical systems, which can be either digital or analog.

\begin{figure}[H]
\centering
\includegraphics[width=\columnwidth]{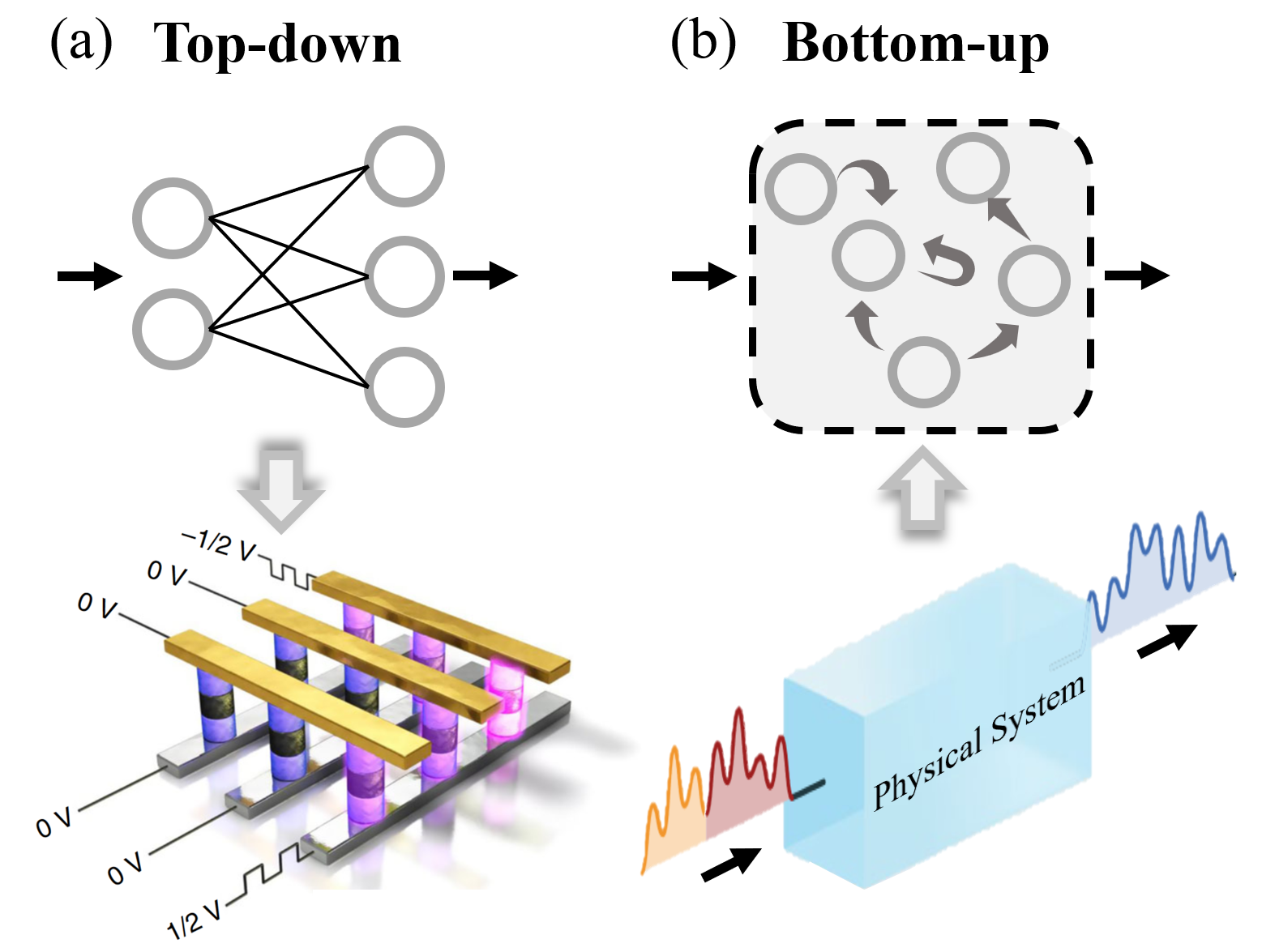}
\caption{Top-down vs bottom-up. (a) Top-down approach maps the computational model of ANN isomorphically into neuromorphic hardwares such as memristor crossbar arrays. (b) Bottom-up approach treats a physical system as a PNN and performs training and inference based on its reconfigurable stimulus-response features. Reprinted with permission from \cite{sun_self-selective_2019,wright_deep_2022}.}
\label{fig:2}
\end{figure}

There are two main approaches to construct PNN as depicted in \cref{fig:2}. The top-down approach involves building and integrating individual physical neurons and synapses based on the blueprint from an ANN. The bottom-up approach entails creating a physical system with a stimulus-response relation that mimics the input-output relation of an artificial neural network, and has adjustable configurations for training.

The top-down approach for constructing PNN typically involves using discrete device elements as synapses, such as interconnecting non-volatile memories \cite{chua_memristor-missing_1971,strukov_missing_2008,burr_neuromorphic_2017,yu_neuro-inspired_2018,ambrogio_equivalent-accuracy_2018}. 
There are emerging memory technologies suitable for emulating a physical synapse \cite{rajendran_neuromorphic_2016}, including metal-oxide-based memristors \cite{jo_nanoscale_2010,mishra_oxygen-migration-based_2019,niu_implementation_2021,rao_thousands_2023}, also known as resistive random access memories (RRAM), spintronic-based memristors \cite{milo_attractor_2017,huang_magnetic_2017,li_magnetic_2017,zhang_spinorbit-torque_2019,song_skyrmion-based_2020,zhang_antiferromagnet-based_2020,fukami_perspective_2018,kurenkov_neuromorphic_2020,lequeux_magnetic_2016}, ferromagnet/ferroelectric tunnel junctions \cite{sengupta_short-term_2016,boyn_learning_2017}, ferroelectric memristors \cite{kim_ferroelectric_2019,ren_associative_2022}, phase change materials (PCM) \cite{shen_deep_2017,cheng_-chip_2017,goi_perspective_2020,sebastian_tutorial_2018,sarwat_phase-change_2022} and organic electrochemical devices \cite{van_de_burgt_non-volatile_2017}.
Implementing PNNs in this way can significantly accelerate inference by taking advantage of the massively parallel operations allowed by physical laws, such as Kirchhoff's law. 
The memristor crossbar array (\cref{fig:1}(d)) is one of the most successful top-down type PNN, as demonstrated by various studies \cite{kim_functional_2012, zhang_neuromorphic_2018, zidan_future_2018, xia_memristive_2019}.  

The bottom-up type PNN, in contrast, is usually based on a physical system that has some specific stimulus-response features. The individual neuron or synapse may or may not be identified. But overall, the system achieves the function of a neural network, which is to perform a nonlinear mapping from input to output, though the mapping of a PNN may not be explicitly generic. For instance, a magnetic thin film with complex magnetization patterns can be used to implement a Hopfield neural network \cite{yu_hopfield_2021}. Random nanocluster networks formed by disordered nanowires and nanoparticles can also exhibit collective intelligence \cite{manning_emergence_2018}. Dynamical systems with high complexity such as coupled oscillators can operate on oscillatory signals, whose synchronization can be used to achieve non-trivial pattern classification tasks \cite{romera_vowel_2018,csaba_coupled_2020}. \add{Emerging devices like dynamic memristors \cite{du_reservoir_2017,zhong_dynamic_2021,zhong_memristor-based_2022}, magnetic tunnel junctions (MTJs) \cite{torrejon_neuromorphic_2017} and material systems such as ferroelectrics \cite{chen_all-ferroelectric_2023,tan_research_2023}, synthetic active particles \cite{wang_harnessing_2024} and artificial spin ice \cite{hu_distinguishing_2023} have been demonstrated to function as a physical reservoir \cite{liang_physical_2024} used in reservoir computing, which utilizes the intrinsic nonlinear dynamics of a given physical system as a computational resource for storage \cite{tanaka_recent_2019}. 

There are trade-offs associated with the top-down and bottom-up approaches for constructing physical neural networks. The top-down approach allows for well-defined network structure and easy implementation of algorithms. However, it faces challenges in precisely replicating idealized network models with real physical devices that have variations and noise. The bottom-up approach  leverages the natural dynamics of the physical platform. However, it is difficult to implement complex architectural designs, and the learning rules may not be fully determined. Overall, top-down PNNs have well-controlled implementations but struggle with non-idealities, while bottom-up approaches are more tolerant of hardware variations but lack programming flexibility. Nevertheless, both approaches can significantly enhance the architecture design of neuromorphic computing. A noteworthy example is the rotating neurons reservoir proposed by Liang \textit{et al.} \cite{liang_rotating_2022}, which adopts a top-down approach by mapping the cyclic reservoir algorithm onto the physical rotation of a dynamic neuron array. This innovative design showcases the potential of integrating algorithmic concepts into physical systems to achieve efficient and effective computation. Conversely, researchers can consider a bottom-up approach by exploring physical systems, such as electronics and mechanical systems, that possess natural rotating components. By embedding and implementing algorithms like the cyclic reservoir within these systems, they can leverage the inherent properties of the physical components to achieve computational capabilities. 
}

\section{Learning and Inference}

Neural network models are effective for two primary reasons. Firstly, they offer enough flexibility to accurately represent real-life situations, with a sufficient number of synapses or weights. Secondly, and more importantly, these models are trainable through efficient algorithms or learning rules. \add{There are two working stages for both ANNs and PNNs, \textit{i.e.,} learning and inference.}

\subsection{Learning for PNN}
Learning or training is the process through which a neural network can adapt itself to the environment, allowing for the construction of a faithful internal representation of the environment \cite{heskes_learning_1991,heskes_learning_1992}. The main objective of the learning process is to identify a set of weights that enable the network to make accurate predictions on new data. The training/learning process can be paced in two steps: firstly, determining how much the weights need to be modified at each learning epoch, and secondly, actually updating the weights stored in the network, either in digital memory or in an analog physical entity.
The learning is achieved by following certain \textit{learning rules}, which govern how the weights should be updated based on the information at hand.
One of the most widely known learning rule is the \textit{Hebbian principle} based on the synaptic plasticity \cite{abbott_synaptic_2000}: \textit{neurons that fire together wire together}, which was first proposed by Donald O. Hebb in his 1949 book ``The Organization of Behavior'' \cite{hebb_organization_2005}. 


Two types of learning are distinguished based on how the learning rules are implemented, as shown in \cref{fig:3}: \add{digital} learning and physical learning. The primary difference between these two types of learning lies in the digitalization of the learning process. \add{Digital} learning implements learning rules in a digital form using algorithms (such as the backpropagation algorithm) running on digital computers, while physical learning embeds the learning rules in an analog form through physical processes within physical systems.





Understanding how to adjust the weight and by how much is crucial part of learning process. \add{Digital} learning uses external training algorithms to calculate this information, while physical learning incorporates this part directly into the physical system. Nowadays, ANNs are primarily trained using \add{digital} learning through training algorithms such as the back-propagation algorithm with gradient descent. However, PNNs can be trained using either \add{digital} learning or physical learning. \add{Digital} learning is more suitable for top-down PNNs, which have well-defined individual neurons and synapses. In contrast, bottom-up PNNs, which lack well-defined weights, are not typically compatible with \add{digital} learning. Instead, their learning process relies on the physical processes inherent in the physical system.

\subsubsection{\add{Digital} learning}

Top-down PNNs, which are designed to accelerate inference, typically lack the ability to adjust independently. Therefore, training of these PNNs must be carried out using \add{digital} learning. In reality, \add{digital} learning of a PNN is quite similar to that of an ANN, with the key distinction lying in the actual update of the trained weight. In ANNs, the weight and its update are digital and stored in computer memories, while in PNNs, the weight and its update are usually carried out through physical control of the analog physical entities that store the weights, or be stored. \add{Recent work on deep PNNs \cite{wright_deep_2022,momeni_backpropagation-free_2023} demonstrate the possibility to store weights (or parameters) digitally, where the PNNs are still controllable by these external parameters.}

Depending on when and how the physical entities are involved in the training process, ideas of online and offline training used in machine learning are borrowed for training PNNs. 
In the context of PNN training, the online and offline training refers to whether the weights are updated physically in an incremental way based on new data or in a single batch process where the weights are calculated based on all training data collected beforehand. Both online and offline training methods have their advantages and disadvantages, and the choice of training method depends on the specific requirements of the neural network and the nature of the data being used for training.

\begin{itemize}
    \item Offline learning, also known as \textit{ex situ} or \textit{in silico} learning, refers to train a PNN completely in computers based on an ANN that is considered as a digital twin of the PNN. Once the ANN has been trained, its weights are injected into its PNN twin. Throughout the whole training process  \cite{jeong_memristors_2016}, the PNN plays no role at all, or regarded as offline. One advantage of offline learning is that synaptic weights can be predetermined using deep learning algorithms on conventional computers. Offline learning has been widely applied in memristor crossbar arrays \cite{kim_functional_2012, zhang_neuromorphic_2018, zidan_future_2018, xia_memristive_2019} for pattern classification \cite{alibart_pattern_2013,bayat_implementation_2018}, images storage \cite{kim_functional_2012} and realization of sparse coding trained by locally competitive algorithm \cite{sheridan_sparse_2017}.  However, there are still challenges that need to be addressed, including material advancements \cite{hoffmann_quantum_2022} and device improvements \cite{tang_bridging_2019}. These challenges involve achieving target conductance values with high accuracy, maintaining low fluctuations and drifts in resistance states, obtaining high absolute resistance values for inference, and ensuring high endurance for repeated programming/training \cite{xia_memristive_2019}.



    \item Online learning, also known as the \textit{in situ} learning, involves updating the network's parameters after each individual data point or small batches of data. This approach is necessary when not all training patterns are available simultaneously \cite{heskes_-line_1993}. The primary advantage of online learning is its ability to adapt to changing environments, where the data distribution gradually shifts over time \cite{murata_-line_2002}. Unlike offline learning, where the weights are determined using all training data in a batch, online learning is a process where the old weight vector $\mathbf{w}$ is iteratively updated to a new one $\mathbf{w}^\prime$. This ability allows the rate of convergence or the degree of accuracy to automatically increase based on whether the weight vector is significantly different from the optimal or nearly optimal solution \cite{amari_theory_1967, heskes_-line_1993, barkai_local_1995}. Although online learning was initially proposed for software implementation, it has provided valuable guidance for enabling physical systems to interact with external circuits, leading to the concept of `on-chip' learning \cite{zhang_sign_2018, zhang_edge_2023}. The online learning strategy has found wide application in RRAM and memristor crossbar arrays for tasks such as digital pattern classification \cite{prezioso_training_2015, li_efficient_2018, zhang_sign_2018, wang_situ_2019}, self-organizing map implementation \cite{wang_implementing_2022}, feature extraction and dimensionality reduction \cite{choi_experimental_2017}, unsupervised K-means clustering \cite{jeong_k-means_2018}, and construction of deep belief neural networks \cite{wang_memristive_2022}. Notably, online learning can also be applied to train spiking neural networks \cite{wang_fully_2018,bellec_solution_2020}, where artificial neurons exhibit leaky integrate-and-fire characteristics, as traditional learning rules for ANNs, such as the conventional backpropagation algorithm, cannot be directly used \cite{wang_supervised_2016}.
\end{itemize}
    


\subsubsection{Physical learning}

Physical learning for PNNs involves integrating knowledge of physical laws and principles into the training process of neural networks. This includes incorporating fundamental laws of physics, such as conservation of energy, momentum, and other physical constraints, into the design and training of the neural network. Depending on the involvement of physics in the learning process, we can distinguish between physics-aware learning, where physical laws play a more passive role, and physical self-learning, where physical laws play an active role. When physical laws actively contribute to learning, the PNN may acquire self-learning capability, similar to various life forms that adjust themselves to better adapt to their environment.

\begin{figure}[H]
\centering
\includegraphics[width=\columnwidth]{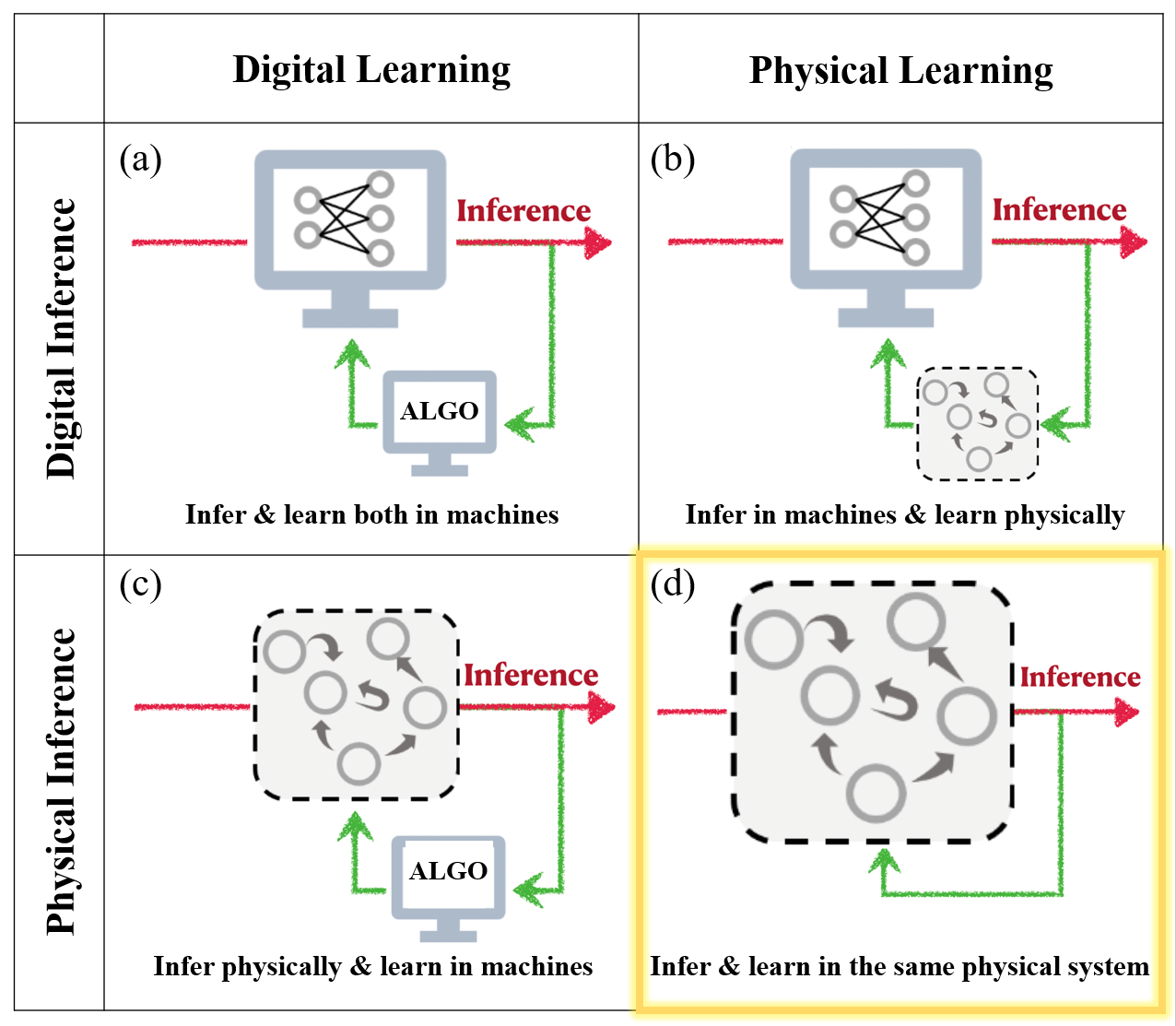}
\caption{\add{Digital} learning vs Physical learning. Neuromorphic computing can be divided into four categories regarding to whether the learning and inference are performed in (a,c) digital machines by algorithms (ALGO) or (b,d) physical systems by embedded physical process. Physical self-learning belongs to the paradigm where both inference and learning are carried out within the same physical system.}
\label{fig:3}
\end{figure}

\begin{itemize}
    \item  Physics-aware learning is a hybrid \textit{in situ}\,-\,\textit{in silico} algorithm to solve the mismatch between the analog physical weight and its digital representation in the machine learning algorithms, first proposed by Wright \textit{et al} \cite{wright_deep_2022}. 
    Physics-aware learning combines the advantages of backpropagation algorithm, the \textit{de facto} training method for large-scale neural networks, with the ability to train controllable physical systems. By breaking the traditional software-hardware division, the physics-aware learning allows for the training of deep PNNs made from layers of controllable physical systems, even when these physical layers lack mathematical isomorphism to conventional artificial neural network layers. This approach enables the scalability of backpropagation while automatically mitigating imperfections and noise achievable with \textit{in situ} algorithms. The concept is universal and has been applied experimentally on optics, mechanics and electronics systems \cite{wright_deep_2022}. Similar concept has been applied in designless nanoparticle network, whose electrical properties depend on input-voltage signals from multiple electrodes. The nanoparticle network can be optimized using a genetic algorithm or a deep-learning approach, aiming to perform reconfigurable computational tasks after training \cite{bose_evolution_2015,ruiz_euler_deep-learning_2020}. 
    The physics-aware learning applies to both top-down and bottom-up PNN. 

    \item 
   Physical Self-Learning, as described by Feldmann \textit{et al.} \cite{feldmann_all-optical_2019}, Yu \textit{et al.} \cite{yu_hopfield_2021} and L\'{o}pez-Pastor \textit{et al.} \cite{lopez-pastor_self-learning_2023}, involves a learning process where the internal degrees of freedom of a PNN evolve in response to a given environment or stimulus based on physical laws enabled by the physics system, rather than relying on external computation. As a result, the weights encoded within these internal degrees of freedom can autonomously adjust themselves through physical evolution without the need for external computations, as depicted in \cref{fig:3}(d). 
   It is important to note that the concept of \textit{physical self-learning} discussed here is synonymous with the notion of \textit{autonomous training} proposed in \cite{buckley_photonic_2023}, and it is closely related to \textit{autonomous physical learning} as introduced in \cite{stern_learning_2023}, albeit with a more specific definition. 
   
    
    
    
\end{itemize}

In order to facilitate the self-learning capability of a PNN, specific requirements must be met by the underlying physical system. The learning process involves feeding back the inference error to the neural network, and a component within the system will utilize this error information to adjust the neural network. In traditional machine learning, this component is the digital computer. This process is to learn from the mistakes, which is to tune the weights to reduce mistakes. Equally efficient, neural network can also learn from correctness, which is to strengthen the weights when correct answer is fed to the network. For self-learning, the weight-tuning must be carried out autonomously by the PNN itself without external intervention. Therefore, such error/correctness sensing should be built-in in the PNN, and the response to the sensed error/correctness must also be embedded in some physical process within the PNN. 

An example of achieving autonomous sensing and updating involves a non-volatile resistor with positive feedback. For instance, the resistance decreases (increases) when a large (small) current flows through it, and this resistance change is maintained when the current stops flowing. This mechanism strengthens the correct flow pattern injected into the network, enabling self-learning. We anticipate that such self-learning capabilities could arise in large complex systems exhibiting emergent collective dynamics \cite{chialvo_emergent_2010}. By harnessing the embedded learning rules within physical systems, the construction of a PNN becomes achievable using a bottom-up methodology \cite{miller_evolution_2002}, as opposed to the conventional top-down design approach used for electric circuits.

\add{\subsection{Inference for PNN}
The inference process differs significantly between traditional digital learning systems and emerging physical learning paradigms. In digital ANNs, inference involves feeding input data forward through the network layers via numeric matrix multiplications calculated step-by-step using computer processors. This digital computation of inferences consumes large amounts of energy and time, particularly for deep networks with many layers \cite{markovic_physics_2020}. In contrast, PNNs as well as hardware implementation of ANNs implement inference directly within analog physical systems using massively parallel natural dynamics. For example, memristor crossbars perform matrix-vector multiplications intrinsically via Kirchhoff's law \cite{jeong_memristors_2016}, enabling ultrafast, low-power inferences by analog signal propagation through interconnected devices. Other approaches use nanophotonic circuits \cite{feldmann_all-optical_2019}, spintronic oscillators \cite{kanao_reservoir_2019,romera_vowel_2018} or elastic/fluidic systems \cite{bhattacharyya_memory_2022,stern_supervised_2021} to infer outputs from optical, magnetic or mechanical inputs. This physical acceleration circumvents the intrinsic inefficiencies of conventional processors,  which serves as a primary motivation driving the field of neuromorphic computing \cite{christensen_2022_2022}.
}

\add{
\subsection{Four scenarios for neuromorphic computing}

We categorize four scenarios for neuromorphic computing in \cref{fig:3}, regarding to whether the learning and inference are performed in conventional computers by algorithms or in physical systems according to physical process.

\begin{itemize}
    \item  Inference and learning are both performed in digital machines (\cref{fig:3}(a)). This aligns with the well-established field acknowledged as ``machine learning'' and falls outside the scope of this review.

    \item Inference performed in digital machines while learning is performed in physical systems (\cref{fig:3}(b)). This is a relatively unexplored scenario, because the advantage of physical inference on physical acceleration was first realized by the community of neuromorphic computing while physical learning is still in its infant stage. However, a rapidly emerging field known as Physics-Informed Neural Networks (PINN) \cite{cuomo_scientific_2022} has gained attention, which approximate solutions to partial differential equations (PDEs) for various domains, including fluidic \cite{cai_physics-informed_2021}, atomistic \cite{pun_physically_2019}, and quantum dynamics \cite{norambuena_physics-informed_2024}. 

    \item  Inference is performed physically and learning is performed in digital machines (\cref{fig:3}(c)). Most neuromorphic hardware, such as memristor crossbar arrays \cite{zhang_neuromorphic_2018,xia_memristive_2019} and other types of PNNs \cite{wright_deep_2022,momeni_backpropagation-free_2023}, align with this scenario.

    \item  Both learning and inference are performed in the same physical system (\cref{fig:3}(d)). In this scenario, physical laws can either passively contribute through physics-aware learning or be actively involved according to physical self-learning. In the subsequent section, we will review recent advancements that have either implemented or hold the potential to achieve the concept of physical self-learning.

\end{itemize}

}

\section{Physical Systems Capable of Physical Self-learning \label{sec:3}}

\subsection{Self-routing in random nanocluster networks \label{sec:3-1}}

\begin{figure}[b]
\centering
\includegraphics[width=1\columnwidth]{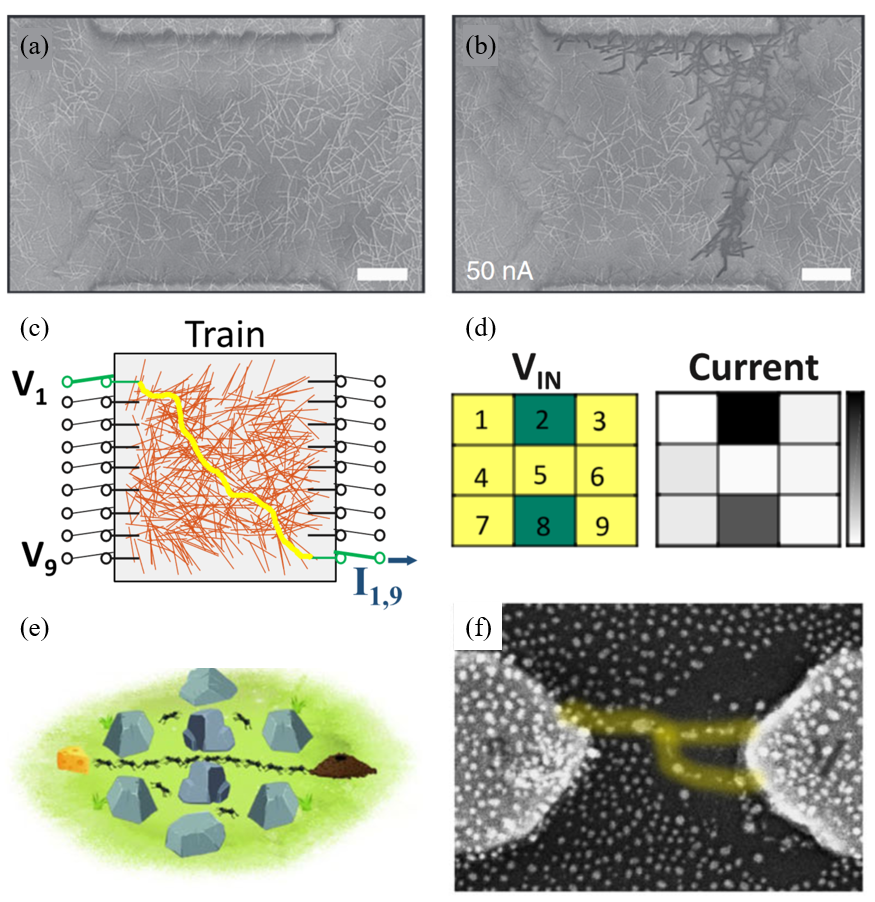}
\caption{A winner-takes-all conducting path is formed, corresponding to the lowest-energy connectivity path (a) before and (b) after the electric field is applied. Reprinted with permission from \cite{manning_emergence_2018}. (c) Interconnected nanowire pathway with low resistance is formed when voltage difference is applied on corresponding electrodes. (d) A 2-bit pattern is stored as an ensemble of higher connectivity pathways between the selected input/output channels. Reprinted with permission from \cite{diaz-alvarez_associative_2020}. (e) Schematics of the ant colony algorithm. (f) Implementation of ant colony algorithm based on Ag nanoclusters, analogous to the ants. Reprinted with permission from \cite{xu_distributed_2022}.}
\label{fig:4}
\end{figure}

 Disordered clusters of nanowires and nanoparticles can form networks that exhibit collective intelligence. Manning \textit{et al.} \cite{manning_emergence_2018} demonstrated the emergence of ``winner-takes-all (WTA)'' paths in macroscale nanowire networks. In WTA algorithms, neurons in a layer, typically the output layer, compete with each other for activation until only one neuron emerges as the winner and becomes active, representing the neuron associated with the strongest input signal. In this case, the WTA behavior is physically manifested through the clustering of mobile ions within the metal-insulator-metal junction, induced by an applied electric field. The largest cluster grows more rapidly than others, as depicted in \cref{fig:4}(a) and (b). The experiments indicate that the network dominated by junctions possesses the capability to autonomously select the lowest energy connectivity pathway within a complex random network. Similar to biological systems, memory is stored in the conductance state and encoded within specific connectivity pathways, which holds significant implications for the application of reservoir computing. A similar functionality can be achieved using molecular neuromorphic networks composed of single-walled carbon nanotubes complexed with polyoxometalate \cite{tanaka_molecular_2018}.

Diaz-Alvarez \textit{et al.} \cite{diaz-alvarez_emergent_2019,diaz-alvarez_associative_2020} designed and demonstrated an associative device based on nanowire networks with resistive switching and retained connectivity. As illustrated in \cref{fig:4}(c), associative routing between input node 1 and output node 9 can be established by reducing the resistance in the nanowire networks through voltage activation. This leads to the formation of multiple pathways with low resistance, thereby enhancing nanowire-to-nanowire connectivity. Subsequent cycles of training and testing strengthen the local connectivity between the selected channels, enabling the simultaneous training of multiple patterns within the same network. A 2-bit pattern, exemplified in \cref{fig:4}(d), can be stored, and the network accurately retrieves the trained pattern during the testing procedure.

Swarm intelligence-inspired algorithms, including genetic algorithms and ant colony algorithms, have played a crucial role in optimization and distributed artificial intelligence. Xu \textit{et al.} \cite{xu_distributed_2022} presented a remarkable demonstration of a multi-agent hardware system that utilizes distributed Ag nanoclusters as physical agents. In response to external electrical stimuli, the Ag nanoclusters adaptively evolved to form different connectivity patterns based on combined input schemes, as illustrated in \cref{fig:4}(e) and (f). The resulting connection path represents the optimal decision, considering both the cost (distance) and reward (voltage). The self-organization of the Ag physical network, driven by positive feedback from information interaction, leads to a significant reduction in computational complexity. The proposed hardware system exhibits efficient solutions for graph problems and optimization tasks, such as demonstrating gradient descent route planning with self-adaptive obstacle avoidance.

\subsection{Memory formation in adaptive elastic networks and flow networks \label{sec:3-2}}

The effect of ``Directed Aging'' has been shown by Pashine \textit{et al.} \cite{pashine_directed_2019} that the elastic properties of materials can depend strongly on the history of the imposed deformations, which obeys a natural greedy algorithm and follows the path of most rapid and accessible relaxation, making elastic networks an intriguing platform to discuss about memory formation. Stern \textit{et al.} \cite{stern_continual_2020} theoretically studied an elastic network whose links grow and strengthen over time according to the local geometry of the network. It is found that such a network can acquire desired stable states in sequence only if the elastic elements are sufficiently nonlinear. The elastic networks is constructed with $N$ nodes, and each two nodes are connected by a spring of stiffness $k_{ij}$ and rest length $l_{ij}$. The position of node $i$ is described by $\mathbf{x}_i$, and the energy of the elastic network is determined by
\begin{equation}
    E(\{\mathbf{x}\})=\frac{1}{2}\sum_{i=1}^{N}\sum_{j=i+1}^{N} k_{ij}\left(r_{ij}-l_{ij}\right)^\xi,
\end{equation}
with $r_{ij}\equiv\|\mathbf{x}_i-\mathbf{x}_j\|$ the distances between nodes and $\xi$ parametrizes the nonlinearity. When a spring is pulled away from its preferred length, it exerts a force given by $F\sim\left(r-l\right)^{\xi-1}$. The behavior of the spring depends on the exponent $\xi$. For $\xi=2$, the spring is linear, meaning that the force increases as the spring is strained further. However, for $\xi<2$, the spring exhibits softer restoring forces at larger distances. In particular, for $\xi<1$, the spring's response weakens as it is strained.

In the elastic network, two learning strategies can be applied: simultaneous stabilization and continual learning. Simultaneous stabilization aims to stabilize multiple desired states by solving the equations $0=\partial_{\mathbf{x}_a}E$ on each node, finding appropriate stiffness values $k_{ij}$ and rest lengths $l_{ij}$. The capacity of such networks to store multiple stable states is expected to scale linearly with the system size. However, adding a new stable state requires a complete rewiring of the network with new springs. In contrast, continual learning, as proposed by Stern \textit{et al.} \cite{stern_continual_2020}, introduces the concept of an effective spring constant that grows with time for the rods connecting two nodes $i$ and $j$, which is given by
\begin{equation}
\frac{dk_{ij}^{\text{eff}}}{dt}=k_0 f(r_{ij}).    
\end{equation}

\begin{figure}[b]
\centering
\includegraphics[width=1\columnwidth]{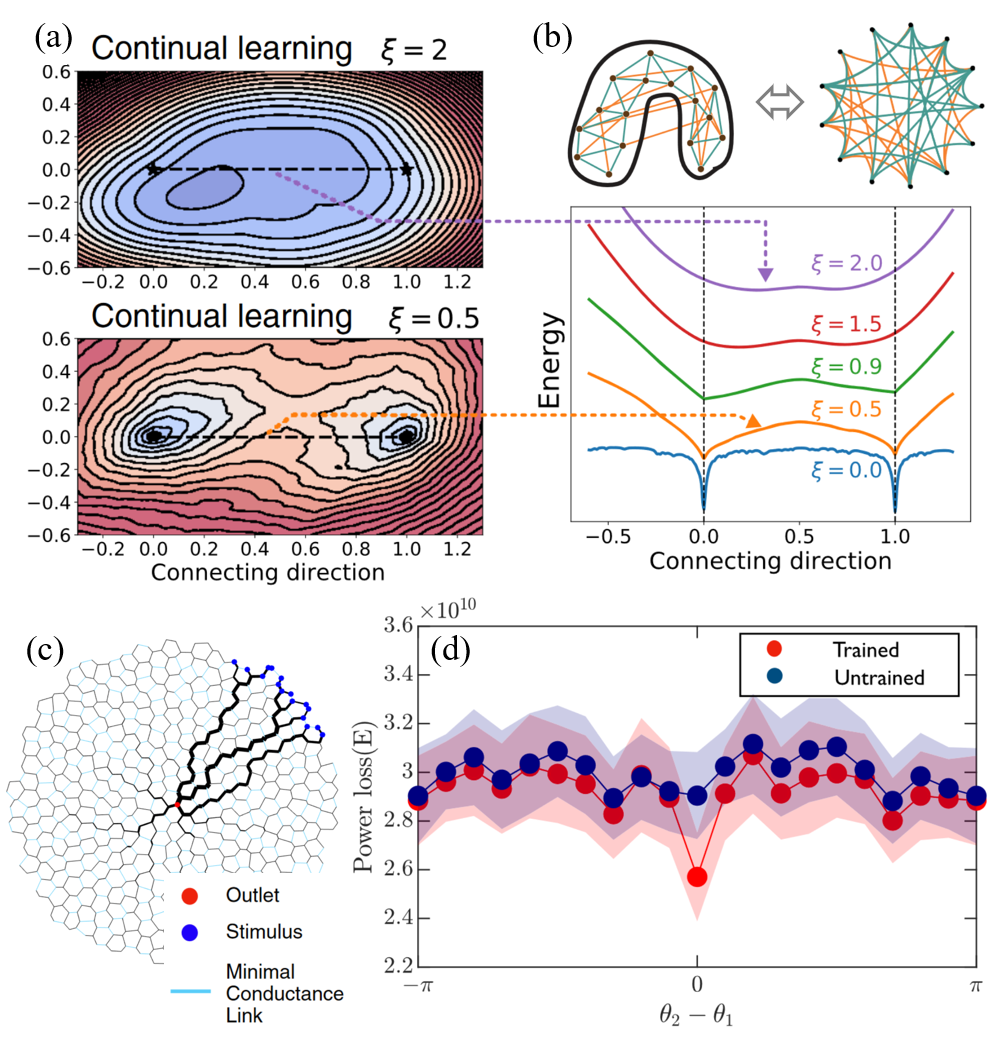}
\caption{(a) Continual learning in elastic networks can learn multiple states according to incremental growth of effective springs, as schematically shown in (b). Reprinted with permission from \cite{stern_continual_2020}. (c) A flow network evolved with a stimulus equally distributed over the blue filled nodes. (d) Power loss before and after training, where training stimulus is applied at the angle $\theta_2=\theta_1$. Reprinted with permission from \cite{bhattacharyya_memory_2022}.}
\label{fig:5}
\end{figure}

Continual learning involves an incremental growth model, where unstretched elastic rods gradually form between every pair of nodes when the material is left in a configuration for a certain period. The growth rate $f(r_{ij})$ depends on the separation between the nodes, as illustrated in \cref{fig:5}(a) and (b). It has been found that continual learning without forgetting requires nonlinear springs. These nonlinearities are essential for a broad class of continual models where the spring network is updated based solely on the current configuration, without knowledge of past or future desired configurations. The frameworks of simultaneous stabilization and continual learning have complementary strengths. The implementation of physical continual learning requires a nonlinear spring while it is possible to apply the concept of simultaneous stabilization applies for spring-node systems with any value of nonlinearity. Continual learning may outperform simultaneous stabilization regarding to quality of encoding (such as barrier height and attractor size).

Similar to the elastic networks, the continuous adaptation of flow networks has been proven to play an important role in memory formation. Bhattacharyya \textit{et al.} \cite{bhattacharyya_memory_2022} theoretically investigated an adaptive flow network and show that adaptation dynamics allow a network to memorize the position of an applied load within its network morphology. As depicted in \cref{fig:5}(c), flow network is a physical model where links are strengthened and weakened over time adaptively. The network is defined as a graph of $N$ nodes connected by links which have length of links $l_{ij}$ and corresponding conductance $C_{ij}(t)$ varying over time $t$. The flow rate can be obtained as $Q_{ij}(t)=C_{ij}(t)(p_i(t)-p_j(t))$, where $p_i$ represents potential at node $i$. The network can be characterized by an energy function (averaging over a period $T$)
\begin{equation}
E(t)=\sum_{\langle ij \rangle}\frac{\langle Q_{ij}(t)^2\rangle_T}{C_{ij}(t)}-\lambda\sum_{\langle ij \rangle}(C_{ij}(t)l_{ij})^\gamma l_{ij},
\end{equation}
where the first term is exactly the power loss and the second term applies a global constraint in the presence of building cost, with $\lambda$ the Lagrange multiplier and the parameter $\gamma$ determining how link conductance contribute to the building cost. An adapting rule on $C_{ij}(t+\delta t)$ can be explicitly derived by solving the Lagrange multiplier. The adaption eventually minimizes the power loss of the whole network given the constraint of material conservation. 

Notably, the parameter $\gamma$ in the building cost has a significant impact on memory formation in flow networks. For instance, when an additional load is applied at a specific angle on the network's boundary over a duration, memory can be formed for $\gamma=0.5$. This process can be considered as a training phase. After training, the power loss is minimized precisely for the angle at which the writing stimulus was applied, indicating that the network configuration has been optimized through memory of the stimulus (see \cref{fig:5}(d)). Memory formation, however, is only possible for $\gamma<1$, while $\gamma=1$ is typical for resistor networks or porous media.

In the field of neuroscience, it has been observed that memory formation is influenced by synaptic plasticity-dependent competition rules \cite{jeong_synaptic_2021}. Biological networks can develop without centralized control, which allows them to yield reasonable solutions to combinatorial optimization problems \cite{tero_rules_2010,kramar_encoding_2021}. This section highlights that both the parameter $\xi$ in elastic networks and the parameter $\gamma$ in flow networks play vital roles in determining the nonlinearity of connections and whether memory formation is possible. Although the discussion focuses on physical networks, it is worth noting that there is a broader scientific community dedicated to understanding memory effects that appear in materials \cite{keim_memory_2019}.

\vspace{1cm}

\subsection{Magnetic textures with intrinsic Hebbian learning as associative memory \label{sec:3-3}}

Energy-based models have garnered \add{interests} from both the physics and machine learning communities \cite{huembeli_physics_2022}. One specific type of artificial neural network is the Hopfield neural network, originally proposed by Hopfield \cite{hopfield_neural_1982}. This network architecture aims to mimic physical systems like the Ising spin-glass model \cite{amit_spin-glass_1985}, exhibiting emergent collective computational abilities. One of the prominent applications of the Hopfield neural network is associative memory. Associative memory is a dynamical system concerned with the storage and retrieval of memories \cite{hoover_memory_2023}. In this framework, uncorrupted memories are associated with local minima in the network's energy landscape. Conversely, more corrupted memories correspond to higher energy values. The Hopfield network is capable of reconstructing an entire memory from partial information through a dynamic inference process, as long as the provided partial information is sufficient to identify a single memory.

\begin{figure}[H]
\centering
\includegraphics[width=1\columnwidth]{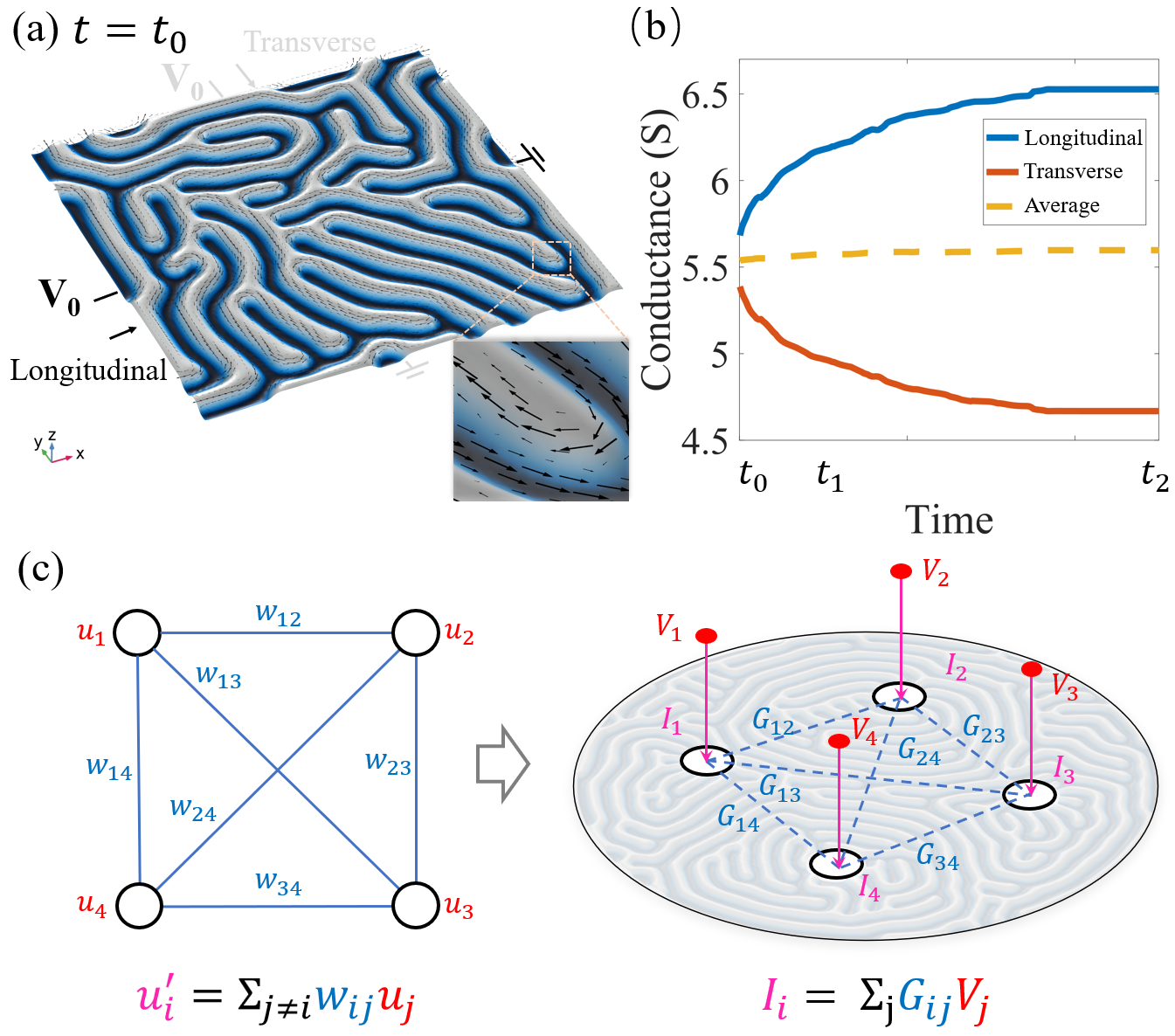}
\caption{(a) Snapshot of a maze domain structure. (b) With applied voltage, current flowing through the magnetic textures exerts spin-transfer torque, so that the spatial distribution of magnetic configuration as well as conductance state evolve with time. (c) The model of a four-node Hopfield neural network is mapped to a magnetic device with four electrodes, with input, output and weight encoded as voltage $V_j$, current $I_i$ and conductance $G_{ij}$ respectively. Reprinted with permission from \cite{yu_hopfield_2021}.}
\label{fig:6}
\end{figure}

Hopfield neural networks are a specific type of recurrent neural network architecture characterized by fully connected neurons. Due to their ease of implementation, these networks have been successfully mapped to various physical systems, enabling the realization of associative memory functions \cite{hu_associative_2015,fukami_magnetization_2016,borders_analogue_2016,fiorelli_signatures_2020,cai_power-efficient_2020,ren_associative_2022,saccone_direct_2022,schnaack_learning_2022}. In most of these works, the focus is primarily on the inference process, which involves retrieving information from the network after the training stage. During training, the weights of the network are computed using algorithms implemented on conventional computers, aligning with the paradigm depicted in \cref{fig:2}(c).

Yu \textit{et al.} \cite{yu_hopfield_2021} proposed a hardware scheme that enables self-learning within the framework of Hopfield neural networks. The physical system they considered consists of conducting magnetic films with inhomogeneous textures, as depicted in \add{\cref{fig:6}}(a). When subjected to electric currents, the magnetic textures undergo spatial reconfigurations due to spin-transfer torque. This results in \add{increment} in conductance along the longitudinal direction (parallel to the current) and a decrease in conductance along the transverse direction (perpendicular to the current). The conductance evolution, illustrated in \add{\cref{fig:6}}(b), exhibits both excitatory and inhibitory features, indicating that the magnetic textures behave like adaptive synapses.

To demonstrate the feasibility of their approach, Yu \textit{et al.} \cite{yu_hopfield_2021} mapped a four-node Hopfield neural network model onto a magnetic film with four electrodes. The synaptic weights were stored as a conductance matrix defined by distributed electric contacts. The relationship between the output current $I_i$ and the input voltage $V_j$ was described by Kirchhoff's law: $I_i = \sum_j G_{ij}V_j$. The energy function of the physical Hopfield neural network can be expressed as
\begin{equation}
    E=-\sum_{i,j}G_{ij}[\mathbf{m}(\mathbf{r})] V_i V_j,
\end{equation}
where $V_i$ is binary voltage applied on each electrode and $G_{ij}[\mathbf{m}(\mathbf{r})]$ is element of conductance matrix which depends on spatial distribution of magnetic textures, which is determined during the self-learning process. Because of the plasticity of the magnetic texture, electric currents can train the network by modulating the texture as well as corresponding conductance (weight) matrix, without interference of external computational algorithms.

In conventional training strategies for Hopfield neural networks, the synaptic weights are determined by computers using algorithms such as the outer product method $w_{ij}=\sum_{s=1}^{k}x_i^s x_j^s$ with $w_{ii}=0$ \cite{baldi_symmetries_1987,tolmachev_new_2020}, where $\mathbf{x}^s$ represents the state vector of the pattern $s$ being memorized. However, the approach proposed by Yu \textit{et al.} \cite{yu_hopfield_2021} differs from conventional implementations. Instead of relying on external computation, the malleable magnetic textures autonomously train the network according to an incremental Hebbian learning principle:

\begin{equation}
G_{ij}^{n+1}=G_{ij}^n+\Delta_{ij}^n[{V_i}].
\end{equation}

A possible explicit form of this learning rule has been recently investigated \footnote{Niu \textit{et al}. Self-learning magnetic textures with intrinsic gradient
descent adaption. Unpublished.}, and it shares similarities with the well-known Oja's rule \cite{oja_simplified_1982} which incorporates a ``forgetting'' term to stop the unconstrained growing of weights guided by original Hebbian's rule. Moreover, this self-learning mechanism aligns conceptually with the synaptic plasticity rule proposed by D. Krotov and J. J. Hopfield \cite{krotov_unsupervised_2019}. The incorporation of synaptic ``plasticity'' and ``competence'' required by Hebb's rule is believed to influence memory formation in neuroscience \cite{miller_role_1994,jeong_synaptic_2021}, and now these principles are intrinsically built into the physical process.

While the self-learning mechanism in malleable magnetic textures shows promise, there are several aspects that require further investigation. For example, it is important to explore whether the memory capacity \cite{amit_storing_1985,mceliece_capacity_1987,fusi_memory_2021} can be further improved using the physical self-learning rule and what is the scaling law of self-learning process on either energy consumption or learning duration when the number of nodes increases. Additionally, investigating the applicability of the self-learning strategy in multilayer networks, such as back-propagation \cite{pineda_generalization_1987}, would be valuable. Recent experimental verification of this concept has been conducted using field-driven permalloy films, where the gradient descent nature of the learning rule has been discovered \cite{niu_self_2023}. Kumar \textit{et al.} \cite{kumar_hybrid_2023} have also demonstrated a similar intrinsic Hebbian rule experimentally using hybrid volatile/nonvolatile memory devices with ternary metal oxide as the active material. The concept is expected to extend to higher dimensions, such as three-dimensional nanoscale gyroid networks \cite{koshikawa_magnetic_2023}.

\subsection{An atomic Boltzmann machine capable of self-adaption \label{sec:3-4}}

Boltzmann machines are powerful tools for solving problems in combinatorial optimization \cite{hopfield_neural_1985,goto_combinatorial_nodate}, but their training algorithm poses challenges and tends to struggle with large problem sizes \cite{patel_logically_2022}. Inspired by such a local learning rule, Kaiser \textit{et al.} \cite{kaiser_probabilistic_2020} designed a probabilistic circuit for autonomous learning, mapping a continuous version of the Boltzmann machine learning rule to a clockless autonomous circuit. While this learning rule can be implemented in a top-down manner, the effective implementation of such a rule in a stochastic physical system remains a challenging question.

\begin{figure}[b]
\centering
\includegraphics[width=0.9\columnwidth]{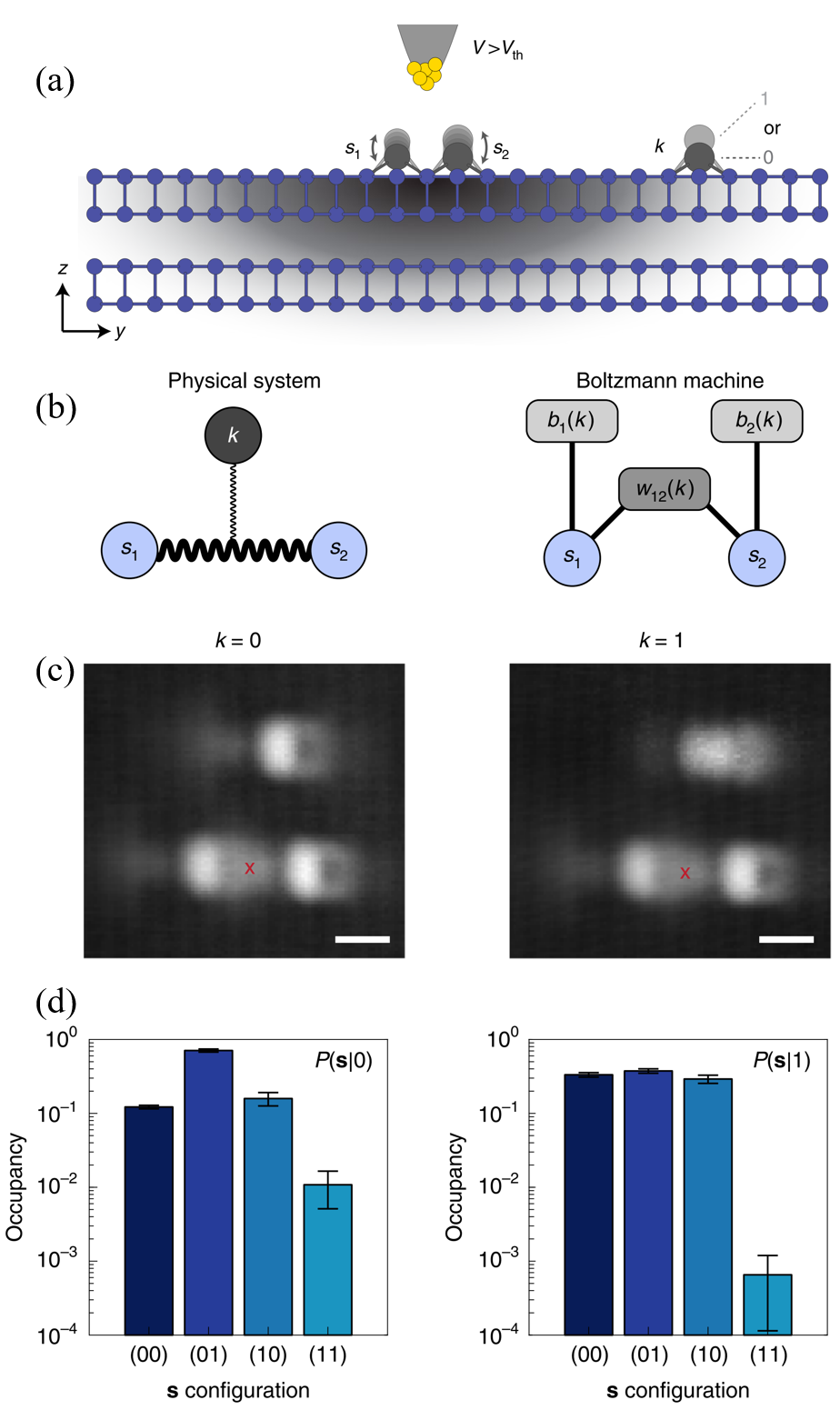}
\caption{Atomic Boltzmann machine with self-adaption. (a) Illustration of physical system consisting of Co atoms on black phosphorus surface, probed by STM technique. (b) Representation of physical system in (a) that mimics Boltzmann machine. (c) STM topography images of two Co atoms $\mathbf{s}$ with a third atom at state $k=0, 1$. (d) Corresponding probability distributions $P(\mathbf{s}|k)$ of two atoms (neurons) $\mathbf{s}$ with $k=0, 1$, respectively. Reprinted with permission from \cite{kiraly_atomic_2021}.
}
\label{fig:7}
\end{figure}

Boltzmann machines are recurrent neural networks that employ stochastic neurons and are capable of solving various challenging computational tasks \cite{aarts_boltzmann_1987,fahimi_combinatorial_2021,fernandez-de-cossio-diaz_disentangling_2023}. Despite sharing a similar structure with Hopfield neural networks, finding an effective training algorithm for Boltzmann machines has been challenging due to their stochastic nature. Ackley \textit{et al.} \cite{ackley_learning_1985} proposed a typical learning rule for Boltzmann machines, where the weight change $\Delta w_{ij}$ is given by $\Delta w_{ij}=\eta(p_{ij}-p_{ij}^\prime)$. Here, $p_{ij}$ represents the average probability of both connected units being in an excited state when the visible units are clamped, $p_{ij}^\prime$ represents the same probability when the network is running freely, and $\eta$ is the learning rate that scales the size of each weight change. Notably, this learning rule is local, relying only on the behavior of the two connected units, without the need for global statistical measures.

In recent research, Kiraly \textit{et al.} \cite{kiraly_atomic_2021} made significant progress in pushing the concept of self-learning to the atomic level by creating a self-adapting Boltzmann machine using cobalt (Co) atoms on black phosphorus. They represented the stochastic neuron physically using the probability distribution of surface Co atoms, where each atom could occupy one of two possible locations, as probed by scanning tunneling microscopy (STM). In their setup, a configuration of three Co atoms was used, with two atoms representing stochastic neurons ($\mathbf{s}$) and the third atom ($k$) influencing the synaptic weight $w_{12}$ through substrate-mediated interactions (\cref{fig:7}(a)). Interestingly, the synaptic weight $w_{12}$ could be tuned in a binary fashion depending on the position of atom $k$ ($k=0$ or $1$), which in turn affected the probability distribution $P(\mathbf{s}|k)$ of the two Co atoms (\add{\cref{fig:7}(b-d)}). Importantly, this self-adapting process occurred without the need for external computational power. The authors also demonstrated the ability to achieve multi-valued synaptic weights using additional Co atoms and the capability to tune the synaptic weights in a semi-continuous manner by adjusting the STM tip voltages. While challenges such as scalability and the underlying learning rule still need to be addressed, this work establishes a promising connection between atomic-level physical interactions and the self-learning capabilities of physical neural networks.

\subsection{Spiking neural networks with STDP learning rule \label{sec:3-5}}

Spiking neural networks (SNN) have emerged as one of the most promising platforms in both the machine learning and material science communities \cite{wang_fully_2018,feldmann_all-optical_2019,zhang_brain-inspired_2020,nguyen_review_2021,han_review_2022,yamazaki_spiking_2022}. Although still in its early stages compared to Deep Neural Networks (DNN), SNN \add{chips} offers ultralow power consumption by encoding and delivering information in spikes, similar to the functioning of the human brain. Notably, SNN chips such as BrainScaleS, TrueNorth, and Loihi have demonstrated remarkable energy efficiency in implementing spiking-based training algorithms through offline or online training approaches \cite{merolla_million_2014,zhang_neuro-inspired_2020}.

\begin{figure}[H]
\centering
\includegraphics[width=1\columnwidth]{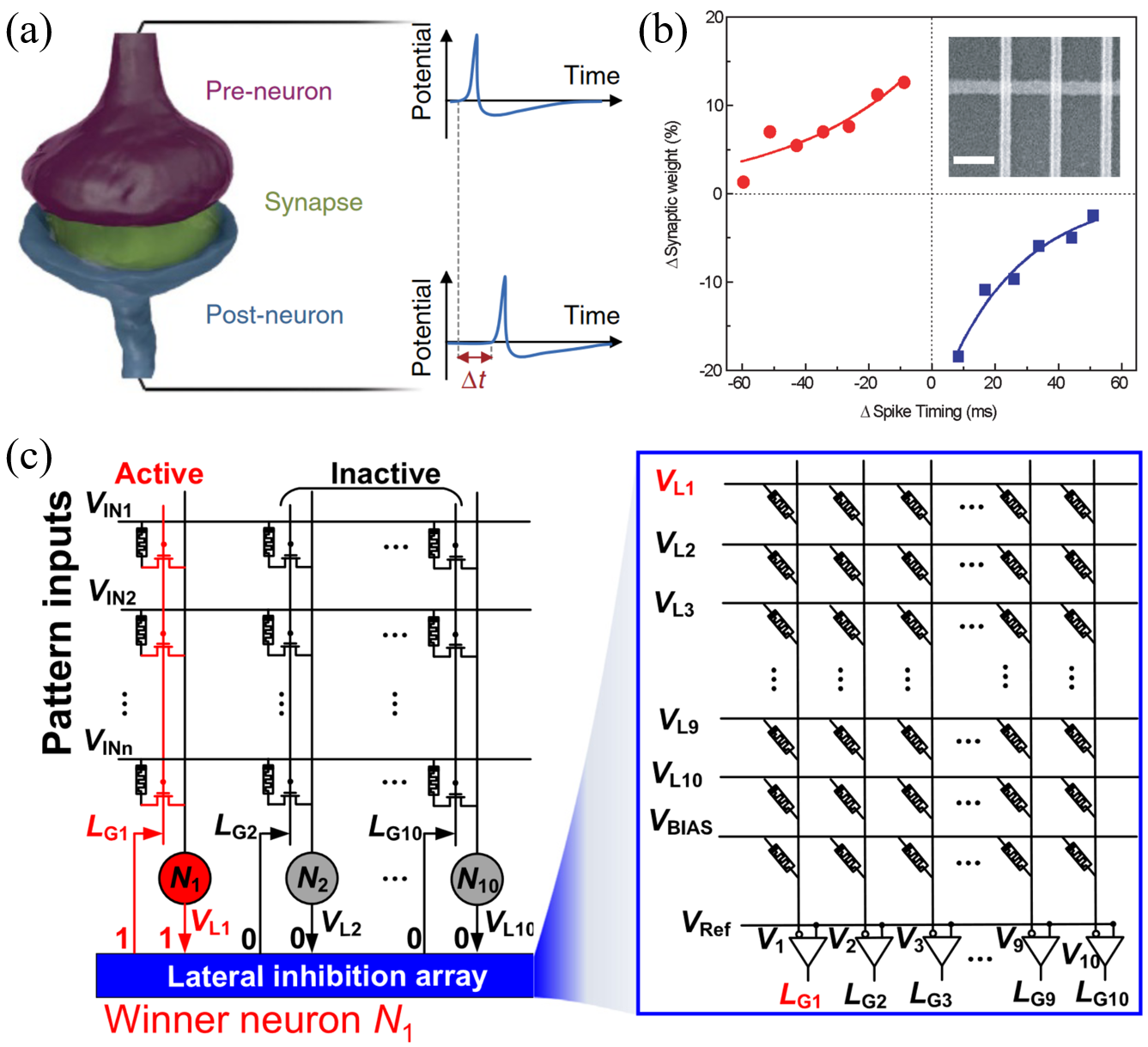}
\caption{Spiking neural networks with STDP for physical learning. (a) Illustration of biological synapses and neurons that encode and pass information with spike signals. Reprinted with permission from \cite{boyn_learning_2017}. (b) STDP behavior in memristive nanodevices. Reprinted with permission from \cite{jo_nanoscale_2010} (c) Schematic of a lateral inhibition array circuit with crossbar array structures. Reprinted with permission from \cite{zhang_hybrid_2021}.
}
\label{fig:8}
\end{figure}

In the realm of learning in physical systems, Spike Timing Dependent Plasticity (STDP) has been extensively studied as an unsupervised learning rule across various material systems. STDP is a synapse-specific Hebbian form of plasticity that implements Hebb's original hypothesis by strengthening synapses when the presynaptic neuron fires before the postsynaptic neuron \cite{song_competitive_2000,jeong_memristors_2016}. This plasticity captures the fundamental workings of biological neurons and synapses, where chemical synapses can be excitatory or inhibitory, promoting or inhibiting the membrane action potential, respectively (\cref{fig:8}(a)) \cite{yamazaki_spiking_2022}. Jo \textit{et al.} \cite{jo_nanoscale_2010} pioneered the implementation of synaptic STDP behavior in memristors, where the synaptic weight is physically represented as electrical conductance in a two-terminal junction, modulated by the difference in pre and post spike timing (\cref{fig:8}(b)) \cite{chua_memristor-missing_1971,strukov_missing_2008}.

The utilization of STDP in physical systems offers several distinct advantages: (1) STDP behavior is applicable to a wide range of materials, including ferroelectric \add{\cite{boyn_learning_2017,yan_ferroelectric_2021}}, magnetic tunnel junctions \add{\cite{borders_integer_2019,sengupta_short-term_2016}}, spin-orbit-torque devices \cite{li_stochastic_2023}, Mott materials \add{\cite{zhang_artificial_2020}}, and optics \add{\cite{feldmann_all-optical_2019}}. This enables collaboration and advancements across multiple research communities, bridging materials science and computational science. (2) Many of these materials are CMOS compatible and can be fabricated as crossbar arrays (\cref{fig:8}(c)), facilitating analog calculations of matrix multiplication directly within the materials. This alleviates the energy consumption typically associated with software-based operations. (3) In SNN, neuronal behavior can be emulated using models such as \add{Leaky Integrate-and-Fire} (LIF), which can be achieved with phase change, magnetic, ferroelectric, and Mott materials. Consequently, the development of fully material-implemented SNN hardware enables highly energy-efficient on-chip learning and self-learning in memristor crossbar arrays \cite{wang_fully_2018,zhao_memristor-based_2020,zhang_hybrid_2021} and all-optical neural networks \cite{feldmann_all-optical_2019}. Ongoing research in this field focuses on improving benchmarks, including linearity, operation voltage, endurance, and other factors, to further enhance energy efficiency and network performance.

Despite the successful integration of STDP and physical devices, this research field is still in its early stages. In terms of SNN, STDP represents a relatively simple learning rule. SNNs, in general, are challenging to train due to the non-differentiable nature of spikes, and it remains uncertain whether SNN performance can match that of DNN-based algorithms. Active research areas include learning mechanisms such as prescribed error sensitivity, ANN-to-SNN conversion, and various forms of STDP \cite{yamazaki_spiking_2022}. The feasibility of implementing these learning mechanisms in materials remains largely unexplored. Additionally, the monotonic increment in memristors may lead to synaptic conductance saturation, which necessitates additional circuit designs to address this issue. At the architectural level, constructing functional autonomous networks with spiking models for relevant tasks is an inspiring research direction \cite{abbott_building_2016}. The exploration of new physical systems with dynamics capable of responding to various network geometries, spike stimuli, and algorithms can introduce novel aspects of efficient learning in materials.

\begin{table*}[t]
\centering
\scriptsize
\begin{tabular}{||c c m{3cm} m{2.5cm} m{2.7cm} m{2.6cm}||} 
 \hline
\textbf{architectures} & \textbf{design approach} & \makecell[c]{\textbf{underlying physical process} \\\textbf{(learning concept)}} & \makecell[c]{\textbf{physical substrate}} & \makecell[c]{\textbf{learning/inference mode}} & \makecell[c]{\textbf{demonstration}}\\ [0.5ex] 
 \hline\hline
\makecell[c]{PNNs with crossbar array structures \\\cite{yi_activity-difference_2023,wang_implementing_2022,kim_functional_2012,zhang_neuromorphic_2018,wang_fully_2018,wang_situ_2019,gao_memristor-based_2022,zhang_edge_2023}} & top-down &  \makecell[c]{digital learning} & \makecell[c]{memristors, ferroelectrics, \\phase change materials, etc.} & \makecell[c]{learn digitally,\\ infer physically} & \makecell[c]{image classification and \\ speech recognition etc.} \\
\makecell[c]{physical reservoirs \\\cite{hu_distinguishing_2023,zhong_dynamic_2021,nomura_reservoir_2021,papp_characterization_2021,kanao_reservoir_2019,pinna_reservoir_2020,chen_all-ferroelectric_2023,wang_harnessing_2024,nakane_reservoir_2018,vidamour_reconfigurable_2023,lee_reservoir_2022,liang_rotating_2022,zhong_memristor-based_2022,liang_physical_2024}} & top-down &  \makecell[c]{digital learning} & \makecell[c]{artificial spin ices, \\dynamic memristors etc.} & \makecell[c]{learn digitally,\\ infer physically} & \makecell[c]{time series prediction and\\spoken-digit recognition etc.} \\
photonic neural networks \cite{pai_experimentally_2023,wang_microring-based_2023,guo_backpropagation_2021,wagner_multilayer_1987} & top-down &  \makecell[c]{backpropagation} & \makecell[c]{photonic chips} & \makecell[c]{learn digitally,\\ infer physically} & \makecell[c]{data/image classification} \\
memristive SNN \cite{zhang_hybrid_2021} & top-down & \makecell[c]{STDP rule} & \makecell[c]{threshold switching and \\resistive switching devices} & \makecell[c]{learn\&infer physically} & \makecell[c]{pattern recognition}\\
all-optical SNN \cite{feldmann_all-optical_2019} & top-down & \makecell[c]{STDP rule} & \makecell[c]{photonic waveguides and \\phase change materials} & \makecell[c]{learn\&infer physically} & \makecell[c]{pattern recognition}\\
deep PNNs \cite{wright_deep_2022,momeni_backpropagation-free_2023} & bottum-up &  \makecell[c]{backpropagation and \\physical local learning} & \makecell[c]{mechanical, optical \\and electronic systems} & \makecell[c]{learn\&infer physically} & \makecell[c]{image classification and\\vowel classification} \\
magnetic Hopfield network \cite{yu_hopfield_2021} & bottom-up & \makecell[c]{intrinsic Hebbian \\(Oja) learning}  & \makecell[c]{magnetic textures} & \makecell[c]{learn\&infer physically} & \makecell[c]{associative memory}\\ 
atomic Boltzmann machine \cite{kiraly_atomic_2021} & bottom-up & \makecell[c]{self-adaptive \\orbital memory} & \makecell[c]{Co atoms on\\ black phosphorus} & \makecell[c]{learn\&infer physically}  &  \makecell[c]{Monte Carlo sampling} \\
nanoparticle networks \cite{bose_evolution_2015} & bottom-up & \makecell[c]{Coulomb blockade and \\single-electron tunneling} & \makecell[c]{Au nanoparticles} & \makecell[c]{learn\&infer physically}  &  \makecell[c]{Boolean logic} \\
nanocluster networks \cite{diaz-alvarez_associative_2020,xu_distributed_2022,manning_emergence_2018} & bottom-up & \makecell[c]{aggregation of nanoclusters \\ with positive feedback} & \makecell[c]{Ag nanowires \\ and nanoparticles} & \makecell[c]{learn\&infer physically}  &  \makecell[c]{gradient descent \\ route planning} \\
nonlinear elastic network \cite{stern_continual_2020,stern_supervised_2021} & bottom-up & \makecell[c]{physical continual learning\\ and physical coupled learning} & \makecell[c]{actin networks, \\kirigami structure etc.} & \makecell[c]{learn\&infer physically} & \makecell[c]{pattern recognition}\\
adaptive flow network \cite{bhattacharyya_memory_2022} & bottom-up  & \makecell[c]{global cost\\ function optimization} & \makecell[c]{vasculatures, \\nonlinear resistors etc.} &  \makecell[c]{learn\&infer physically} & \makecell[c]{memory encoding}\\[1ex] 
 \hline
\end{tabular}
\caption{Some typical neuromorphic architectures and their features.}
\label{table:1}
\end{table*}

\section{Learning Strategies towards Physical Self-learning \label{sec:4}}

While the concept of physical self-learning is still emerging, significant efforts have been dedicated to achieving efficient learning in PNNs. In this regard, various learning strategies have been explored on analog hardware as concluded in \cref{table:1}, moving away from digital approaches. These strategies leverage the intrinsic dynamics of physical systems and aim to minimize reliance on external computation, thereby paving the way for the implementation of physical learning. Although it remains uncertain whether these strategies fully achieve the goal of physical self-learning, they undoubtedly contribute to the advancement of this field.

\begin{itemize}
\item  \textit{Incremental Hebbian learning}. Taking Hopfield network as an example, the incremental Hebbian learning was applied by Hopfield \cite{hopfield_neural_1982} to learn a new memory $\mathbf{x}^s$ and the increment of weight matrix elements takes the form as
    \begin{equation}
        \Delta w_{ij}\propto x_i^s x_j^s. \qquad\text{(incremental Hebbian learning)}
    \end{equation}
    In order to enhance the performance of the network in accessing real memories and in minimizing spurious ones, Hopfield \textit{et al}. \cite{hopfield_unlearning_1983} further proposed an unlearning strategy, where the final equilibrium state $\mathbf{x}^f$ is weakly unlearned by the incremental change 
    \begin{equation}
    \Delta w_{ij}\propto -\epsilon x_i^f x_j^f, \qquad\text{(unlearning)}
    \end{equation}  
    with $0<\epsilon \ll 1$, so that the accessibility of stored memories can be made more nearly even. Other strategies have been proposed such as Grossberg's rule \cite{grossberg_learning_1969} and Oja's rule which considered ``forgetting'' mechanism by adding a contraint term into the learning rule \cite{oja_simplified_1982}. Krotov and Hopfield \cite{krotov_unsupervised_2019} extended the Oja rule to a biologically inspired learning algorithm, called synaptic plastic rule which considers competition between hidden units and implements the Hebbian-like plasticity subjet to homostatic constraints on the length of synaptic weights. The strategies related to incremental Hebbian learning have been widely applied in top-down approach described in \cref{fig:2}. Regarding to physical self-learning, one straightforward way is to look for a non-volatile physical quantity as weights $w_{ij}$ (\textit{e.g.}, conductance or resistance), whose increment is a function of input physical quantities $\mathbf{x}$ (\textit{e.g.}, voltages) as implemented by \cite{yu_hopfield_2021}.

\item \textit{Contrastive Hebbian learning}. The strategy is a generalization of the incremental Hebbian learning including unlearning process mentioned above, which updates the weights proportionally to the difference in the cross products of activations in a clamped and a free running phase which can be applied in continuous Hopfield network \cite{movellan_contrastive_1991} and Boltzmann machines \cite{galland_deterministic_1991}. The key idea of contrastive learning is to introduce a constrastive function $J=\bar{E}^C-\bar{E}^F$ which is difference of energy functions at equilibrium (denoted by ``$\bar{\quad}$'') when the inputs and outputs are both clamped (C), and when the inputs are clamped and outputs are free (F), respectively. Therefore, the system performs the learning successfully when $J=0$. By differentiating the contrastive function $\partial J/\partial w_{ij}$, one can obtain the contrastive learning rule 
\begin{equation}
\Delta w_{ij}\propto \bar{x}_i^{C}\bar{x}_j^{C}-\bar{x}_i^{F}\bar{x}_j^{F}, \qquad\text{(contrastive Hebbian learning)}
\end{equation}
with $\bar{\mathbf{x}}^{C}$ and $\bar{\mathbf{x}}^{F}$ the state vector at equilibrium for clamped phase and freely running phase. After each learning step, the difference in energy at equilibrium between the clamped and free states becomes smaller. It needs to be remarked that the contrastive Hebbian learning is equivalent to backpropagation learning when there are no hidden units \cite{movellan_contrastive_1991}, \textit{i.e.},
\begin{equation}
    \Delta w_{ij}\propto I_i f^\prime(x_j)(t_j-x_j), \ \  \text{(backpropagation, no hidden units)}
\end{equation}
with $t_j$ the teacher vector at outputs, $I_i$ the input vector and $f^\prime$ the derivative of activation function. Backpropagation is one of the most successful algorithms to train neural networks \cite{rojas_neural_2013}. However, it is considered neither biologically nor physically plausible \cite{lillicrap_backpropagation_2020} due to its requirement of a special computational circuit during training, requirement of accurate knowledge about the whole physical system \cite{nakajima_physical_2022} and the need for analytically calculating the derivative of the activation function \cite{hertz_nonlinear_1997}.

\item \textit{Equilibrium propagation}. Equilibrium propagation is a learning framework for energy-based models, sharing similarities with contrastive Hebbian learning and backpropagation, but requiring only one computational circuit and one type of computation for training \cite{scellier_equilibrium_2017}. Equilibrium propagation involves two phases: (1) In the first phase, the inputs are weakly clamped, and the network relaxes to a fixed point. Predictions are then read out from the output units. (2) In the second phase, the outputs are nudged toward their targets, and the network relaxes to a new fixed point corresponding to a smaller prediction error. The system learns the second-phase fixed point by reducing its energy while unlearning the first-phase fixed point by increasing its energy.

Equilibrium propagation differs from contrastive Hebbian learning in certain aspects. The learning rule for equilibrium propagation can be expressed \add{as}
\begin{equation}
\Delta w_{ij}\propto \frac{1}{\beta}(x_i^\beta x_j^\beta-x_i^0 x_j^0), \qquad\text{(equilibrium propagation)}
\end{equation}
where $\mathbf{x}^0$ is state vector at freely running phase (equivalent to $\mathbf{x}^{F}$ in the case of contrastive Hebbian learning) and $\mathbf{x}^\beta$ is the state vector at weakly clamped phase with $\beta$ the clamping factor that controls whether the output is pushed toward the target or not and by how much \cite{scellier_equilibrium_2017}. When $\beta\rightarrow+\infty$, the learning rule comes to the strong clamped scenario as contrastive Hebbian learning ($\mathbf{x}^\beta\rightarrow\mathbf{x}^C$). The concept of equilibrium propagation has found applications in training electric resistor networks \cite{kendall_training_2020, laborieux_scaling_2020} and an Ising machine \cite{laydevant_training_2023}. Additionally, there are other strategies closely related to equilibrium propagation. One such strategy is \textit{contrastive divergence}, which is used for training Boltzmann machines \cite{hinton_training_2002}. Another related strategy is \textit{recurrent backpropagation} \cite{pineda_generalization_1987}.

\item \textit{Physical coupled learning}. Inspired by the concept of \textit{contrastive Hebbian learning} and \textit{equilibrium propagation}, Stern \textit{et al.} \cite{stern_supervised_2021} proposed a general and physically plausible framework called ``coupled local supervised learning'' which has two phases, namely free phase and clamped phase. Given any physical network with a known physical cost function $E(\mathbf{x},\mathbf{w})$, one can drive the relevant coupled learning rule directly according to 
\begin{equation}
    \Delta w_{ij} \propto -\partial_{w_{ij}}(E^F-E^C), \quad\text{(physical coupled learning)}
\end{equation}
where the weights $w_{ij}$ are physical learning degrees of freedom such as pipe conductance in flow networks and spring constant in elastic networks. Because the cost function $E$ can be written as a sum over individual costs of edges $E=\sum_iE_i$, the learning rule is local, \textit{i.e.}, each edge independently adjusts its learning degrees of freedom taking into account only its own response to the applied boundary conditions. Such locality of learning rules promises almost same amount of learning time when applied in networks with larger size. As a proof of principle, the strategy of coupled learning has been experimentally applied to electric resistor networks based on variable resistors \cite{wycoff_desynchronous_2022, stern_physical_2022, dillavou_demonstration_2022}. In these cases, a circuitry is required to locally compare the response of two identical networks subjected to different sets of boundary conditions. The effect of directed aging demonstrated by Pashine \textit{et al}. \cite{pashine_directed_2019} can be seen as a simplified form of coupled learning in the absence of the free term. Anisetti et al. \cite{anisetti_learning_2023} considered a flow network where chemicals can spread via a diffusion process, which can increase or decrease the conductance of pipelines. This approach allows the learning rule to compute gradients using local information for each weight without the need to store the free state. The work by Anisetti \textit{et al}. \cite{anisetti_learning_2023} provides inspiration for conducting coupled learning using two independent physical quantities, which ensures non-interference between the input signal and the feedback signal.

\item \textit{Physical local learning}. Momeni \textit{et al}. \cite{momeni_backpropagation-free_2023} proposed a simple deep neural network architecture augmented by a physical local learning (PhyLL), which enables one to perform supervised and unsupervised learning on a deep PNN without detailed knowledge of the nonlinear physical layer's properties. The strategy is closely related to the \textit{physics-aware learning} \cite{wright_deep_2022}, which aims to train a deep PNN by performing backpropagation on a twin digital model. The improvement of \textit{physical local learning} is to replace the standard backward pass (performed by a digital computer) with an additional single forward pass through a physical system, eliminating the twin modelling phase. Instead of a forward and backward pass applied by standard backpropagation method, PhyLL shares similarity with the concept of ``contrastive learning'' by using two physical forward passes: a positive and a negative forward pass through the physical system, each running on different physical inputs. In each passes, the datasets as well as the correct (incorrect) labels are input together into the multiplayer PNNs and the output on each layer can be expressed as 
\begin{equation}
\mathbf{y}^{\text{pos/neg}}=\mathbf{w}_\text{t}\cdot\mathcal{F}(\mathbf{w}_\text{p}\cdot\mathbf{x}^\text{pos/neg}),
\end{equation}
where $\mathbf{w}_\text{p}$ is untrainable weight matrix provided by physical interconnections and $\mathbf{w}_\text{t}$ is trainable weight matrix to be updated by minimizing a loss function
\begin{equation}
L=\log\left(1+\exp\left(\theta\left(\cos_\text{sim}(\mathbf{y}^\text{pos},\mathbf{y}^\text{neg})\right)\right)\right),
\end{equation}
with $\theta$ a scale factor and $\cos_\text{sim}$ denotes the cosine similarity defined as the cosine of the angle between the two arguments. One can obtain the learning rule according to
\begin{equation}
\Delta \mathbf{w}_\text{t}\propto -\frac{\partial L}{\partial \mathbf{w}_\text{t}}. \qquad \qquad\text{(physical local learning)}
\end{equation}
During the training of deep neural networks using backpropagation, three primary steps must be accomplished: (1) forward propagation, (2) error backpropagation, and (3) gradient computation (updating) steps. PhyLL demonstrates the capability to execute the initial two steps utilizing physical systems, leading to a notable 2/3 enhancement in both speed and energy efficiency when compared to fully digital learning \cite{momeni_backpropagation-free_2023}. In contrast, \textit{physics-aware learning} achieves an approximate 1/3 improvement and the ultimate goal of \textit{physical self-learning} is to accomplish a perfect 3/3, whereby every step of the learning process occurs within physical systems.

\item \textit{Hamiltonian echo backpropagation}. L\'{o}pez-Pastor and Marquardt introduced a general scheme for self-learning in any time-reversible Hamiltonian system \cite{lopez-pastor_self-learning_2023}. Although the idea of physical backpropagation has been proposed in optical neural networks decades ago \cite{wagner_multilayer_1987} and more novel implementations have been realized thereafter \cite{skinner_neural_1995,hughes_training_2018,guo_backpropagation_2021,wang_microring-based_2023,pai_experimentally_2023}, all presently existing proposals are suitable only for some setups with very specific nonlinearities. According to L\'{o}pez-Pastor and Marquardt, the self-learning approach consists of three main steps: (1) Forward pass: The evaluation field $\Psi(-T)$ and the learning field $\Theta(-T)$ (which can represent any physical degrees of freedom in the Hamiltonian system) are prepared at time $t=-T$ and evolve according to Hamilton's equations for a given Hamiltonian. (2) At time $t=0$, a perturbed echo is injected, and a time-reversal operation is performed. (3) Backward pass: The learning field continues to evolve until time $t=T$, and the final result of the entire process is equivalent to the following update of the learning field:
\begin{equation}
    \Theta(T)=\Theta(-T)-i\epsilon\frac{\partial C}{\partial \Theta^*(-T)}, \quad\text{(echo backpropagation)}
\end{equation}
where $C$ is the sample-specific cost function, $\Theta^*$ is the conjugate of learning field, and $\epsilon$ can be understood as the magnitude of variation as a perturbation travelling on top of the time-reversed field from output to input. This self-learning approach requires the physical system to be Hamiltonian and obey time-reversal symmetry. Examples of suitable physical platforms include, but are not limited to, nonlinear nanophotonic circuits \add{\cite{shen_deep_2017,ashtiani_-chip_2022}}, superconducting microwave cavities \add{\cite{blais_cavity_2004,wallraff_strong_2004,majer_coupling_2007}}, cold atoms trapped in optical lattices \add{\cite{PhysRevLett.102.135302}}, and spin wave systems \add{\cite{lan_spin-wave_2015,yu_magnetic_2021,chumak_magnon_2015}} without broken time-reversal symmetry.

\end{itemize}

\section{Discussion and Outlook}

In the last section, we discuss distinctive features, challenges, and opportunities associated with the concept of physical self-learning.

\textit{Common attributes required for self-learning}. As discussed in Section \ref{sec:3}, several common attributes are observed across various physical systems. To emulate Hebbian's principle, a positive feedback process is invariably necessary, which can be commonly observed in a nonlinear conducting system whose conductance strengthens with the flow of current, as introduced in Section \ref{sec:3-1}. However, a monotonous increase in synaptic weight is not universally feasible, either biologically or physically. This typically necessitates a global constraint on weight evolution, and the specific form of this constraint is closely linked to the nonlinear characteristics of physical systems (Sections \ref{sec:3-2} and \ref{sec:3-3}). Moreover, the physical quantities employed to encode weights need not be static and deterministic; they can be dynamic and stochastic, even in the absence of an explicit form (Sections \ref{sec:3-4} and \ref{sec:3-5}). As discussed in Section \ref{sec:4}, the progression of learning strategies in PNNs begins with early versions such as incremental Hebbian learning (learning rule as a function of training data) and advances towards later versions like physical coupled learning (learning rule as a function of feedback from the physical systems). This shift signifies a departure from solely relying on training data and instead harnesses the rich dynamics and interactions within the physical domain.

\textit{Perfection vs imperfection}. Modern architectures for neuromorphic computing, such as memristor crossbar arrays \cite{sheridan_sparse_2017,zidan_general_2018,sun_solving_2019}, strive for perfection. In these architectures, the variation of devices or computing units (such as memristors) is expected to be minimal, as training strategies assume ideal circumstances. However, a gap exists between the perfect neural network model and the imperfect realistic devices. This discrepancy poses challenges, as widely used backpropagation training algorithms for offline and online learning are generally incompatible with hardware due to mismatches between analytically calculated training information and the imprecision of actual analogue devices. To address this issue, training strategies such as the activity-difference technique \cite{yi_activity-difference_2023} and physics-aware training \cite{wright_deep_2022} have been developed. On the other hand, the autonomous learning process in PNNs appears to be more tolerant of imperfections. Strategies like \textit{coupled learning} and \textit{Hamiltonian echo backpropagation} are less sensitive to device variability and can effectively handle noise \cite{lopez-pastor_self-learning_2023}.

\textit{Nonlocal rules vs local rules}. Conventional learning rules implemented on digital computers are generally nonlocal, obtained by performing gradient descent on a global cost function. Consequently, the weight updates have a complex nonlocal dependence on other neurons in the network, making it challenging to implement traditional strategies like backpropagation in PNNs \cite{lopez-pastor_self-learning_2023,stern_supervised_2021}. On the other hand, the physical learning strategies discussed in this review are all local. This means that the weight update between two neurons depends solely on the state of those two neurons and is independent of other neurons in the network. As a result, these local rules enhance the physical plausibility of PNNs, and the weight updates can be performed in parallel, leading to promising scalability of learning speed with increasing network size. While simple physical networks with local learning rules are sufficient for achieving physical self-learning, it remains an open question whether nonlocal learning rules exist in physical systems and whether they can improve learning performance.

\textit{\add{Potential applications}}. In this review, we primarily focus on the implementation of physical (self-)learning in typical structures such as Artificial Neural Networks (ANNs), Spiking Neural Networks (SNNs), and recurrent Hopfield networks. However, extending the concept of self-learning to other structures presents intriguing opportunities. For instance, non-volatile memories acting as artificial synapses can be replaced by wireless connections \cite{ross_multilayer_2023}, and artificial neurons with spiking dynamics can be substituted with nonlinear oscillators \cite{csaba_coupled_2020}. This allows for the construction of Cellular Neural Networks (CNNs) where units or cells communicate with spatially dependent connections \cite{chua_cellular_1988}. The concept of Kohonen neural networks can be applied by considering short-range lateral feedback between neighboring neurons. This enables self-organizing mapping from multidimensional data to lower-dimensional representations \cite{kohonen_self-organized_1982}. Such self-adaptive or self-organized behaviors are expected to be prevalent in physical systems. Constructing continuous attractor neural networks can predict the trajectories of moving objects without the need for training \cite{zheng_anticipative_2020}. It is anticipated that more novel functions can be realized by applying physical self-learning to dynamical synapses. PNNs with self-learning capabilities can be combined with front-end devices to perform multimodal in-sensor computing \cite{zhou_near-sensor_2020}. For instance, artificial retinas can be designed by connecting self-adaptive PNNs with photoelectric receivers \cite{feng_retinomorphic_nodate,liao_bioinspired_2022}. Lastly, PNNs can enter the quantum regime by leveraging the infinite degrees of freedom in phase space \cite{markovic_quantum_2020,labay-mora_quantum_2023}.

\add{\textit{Challenges and opportunities}. While significant advances have been made in realizing physical self-learning, both challenges and opportunities remain for this promising field. One of the key obstacles lies in enhancing device properties, such as nonlinear conductance thresholds, endurance, and variability tolerance. Additionally, there are unresolved architectural questions pertaining to the implementation of diverse network structures and complex algorithms within physical systems. At a systems level, comprehending emergent behaviors arising from collective physical dynamics poses analytical difficulties. From a materials standpoint, there are exciting prospects in identifying new platforms with embedded intelligence, such as 2D \cite{zhu_hybrid_2023,pan_2d_2023} or spintronic \cite{grollier_neuromorphic_2020} systems. Boosting capabilities can be achieved by developing learning rules inspired by life-like principles of homeostasis, competition, and memory formation. Moreover, integrating PNNs with sensing and actuation capabilities empowers potential applications like near-sensor computing \cite{zhou_near-sensor_2020}, artificial retinas \cite{mahowald_silicon_1991} and autonomous robotics \cite{pei_towards_2019}. By addressing these challenges through multidisciplinary collaboration, there is the potential to revolutionize computing using intelligent matter programmed by the laws of physics.
}

\textit{Conclusion.}
In this review, we have explored the concept of physical learning, particularly physical self-learning, which represents a promising frontier in the field of neuromorphic computing. Our investigation suggests that physical systems have the potential to exhibit intelligence by interacting with the environment, receiving and responding to external stimuli, and internally adapting their structures to facilitate the distribution and storage of information \cite{kaspar_rise_2021}. Generally speaking, the concept of physical self-learning can be understood as an optimization process performed by physical optimization machines, which evolve based on the principle such as least action and least power dissipation \cite{vadlamani_physics_2020}. By delving deeper into the concept of self-learning, we aim to gain a better understanding of natural intelligence \cite{callatay_natural_2014,van_gerven_computational_2017}. Achieving this goal requires collaboration among various scientific and engineering disciplines, as they contribute to the exploration and advancement of self-learning.

\section{Acknowledgements}
This work was supported by National Key Research Program of China (2022YFA1403300, 2020YFA0309100), National Natural Science Foundation of China (12204107, 12074073), the Shanghai Municipal Science and Technology Major Project (2019SHZDZX01), Shanghai Pujiang Program (21PJ1401500) and Shanghai Science and Technology Committee (21JC1406200, 20JC1415900). Weichao Yu acknowledges fruitful discussion with Huanyu Zhang, Chuanlong Xu and Lin Xiong.

\end{document}